\documentstyle[prc,aps,epsf]{revtex}
\newcommand{\nsigma}{\mbox{\boldmath $\sigma$}}
\newcommand{\ntau}{\mbox{\boldmath $\tau$}}
\newcommand{\half}{\frac{1}{2}}
\newcommand{\br}{{\bf r}}
\newcommand{\la}{\mbox{$\Lambda$}}
\newcommand{\what}[1]{\widehat #1}
\begin{document}
\title { {\bf A correlated model for $\Lambda$--hypernuclei} }
\author{
F. Arias de Saavedra$^{1)}$, 
G. Co'$^{\,\, 2)}$, and 
A. Fabrocini$^{3)}$
}
\address{
$^1)$ Departamento de F\'{\i}sica Moderna,
Universidad de Granada, E-18071 Granada, Spain \protect\\
$^2)$ Istituto Nazionale di Fisica Nucleare,
Dipartimento di Fisica, Universit\`a di Lecce,
I-73100 Lecce, Italy \protect\\
$^3)$ Istituto Nazionale di Fisica Nucleare,
Dipartimento di Fisica, Universit\`a di Pisa,
I-56100 Pisa, Italy 
}
\maketitle
\date{\mbox{ }}
\vskip 1.0 cm
\begin{abstract}

We study the properties of hypernuclei containing one 
$\la$ hyperon in the framework of the correlated basis function theory
with Jastrow correlations. Fermi hypernetted chain integral equations 
are derived and used to evaluate energies and one--body densities 
of $\la$--hypernuclei having a doubly closed shell nucleonic core in the $jj$ 
coupling scheme, from Carbon to Lead. We also study hypernuclei having 
the least bound neutron substituted by the $\la$ particle. 
The semi-realistic Afnan and Tang nucleon-nucleon
potential and Bodmer and Usmani $\la$-nucleon potential are adopted. 
The effect of many--body forces are considered by means either of a 
three body $\la$-nucleon-nucleon potential of the Argonne type or of a
density  dependent modification of the $\la$-nucleon interaction,
fitted to reproduce the $\la$ binding energy in nuclear matter. 
While Jastrow correlations underestimate 
the attractive contribution of the three body $\la$ interaction, 
the density dependent potential provides a good description of the
$\la$ binding energies over all the nuclear masses range, 
in spite of the relative simplicity of the model.
\end{abstract}
\vskip 1.cm
\pacs{21.80.+a, 21.10.Dr, 21.60.Gx}

%
%

\section{Introduction}
\label{intro}
The microscopic study of atomic nuclei has gained in the recent years 
a noticeable level of reliability, following the improved 
knowledge of the interaction between the nucleons and 
the development of new and more  sophisticated 
many--body theories. It is now possible to exactly solve the 
Schr\"odinger equation for nuclei containing up to $A=8$  
nucleons\cite{wir00}, while the structure of medium and heavy nuclei can 
be to a large extent understood by means either of variational 
techniques\cite{fab00} or of highly refined perturbative 
expansions\cite{hei99}. These relatively recent developments 
have clearly stressed several inadequacies of the independent particle 
model (IPM), on which the time honoured shell model is based. 
For instance, the IPM completely fails in describing the large energy 
behaviour of the one nucleon momentum distribution and the quenching 
of the spectroscopic factors\cite{fab01}. 

The scenario is not equally bright for hypernuclei, or bound systems of 
nucleons and one or more strange baryons, as the $\Lambda$ or $\Sigma$ 
hyperons. The many--body  techniques developed for the nuclei can be, in 
most of the cases, straightforwardly extended to the hypernuclei, at least from 
a formal point of view. However, the weak point of a consistent 
microscopic approach now lies in the poorer understanding of the 
hyperon--nucleon (YN) and hyperon--hyperon (YY) interactions with respect to 
the nuclear case. The limited amount of information on 
the YN scattering data is a consequence of the experimental difficulties 
due to the short lifetime of the hyperons and to the low intensity of 
the beam fluxes. For this reason phenomenological models and effective 
YN interactions have been often used in the study of hypernuclei. 
Approaches based upon Skyrme interactions \cite{ray76} and Woods--Saxon
\cite{mil88} potentials as well as relativistic mean field
theories \cite{sch92,ma96,tsu97} have been used.
However, also microscopic YN potentials have been recently 
produced \cite{mae89,hol89,rij99} and adopted to study the structure of 
hypernuclei from $^{17}_{~\Lambda}$O to 
$^{208}_{~~\Lambda}$Pb \cite{vid98,cug00}.
The $\la$ single--particle properties have been investigated in details
in Ref. \cite{usm99}, making use of the nuclear matter Fermi hypernetted 
chain results in a local density approximation.

In this paper we present a microscopic study of the structure of
various hypernuclei containing a single $\Lambda$ hyperon. This study
has been done by extending the correlated basis functions theory and the  
Fermi hypernetted chain (FHNC) equations technique \cite{ros82} to
describe nucleonic systems containing one hyperonic impurity.
The nucleon--nucleon (NN) interaction we have chosen is
the semi-realistic two--nucleon S3 potential of Afnan and Tang
\cite{afn68}, as modified in Ref.\cite{gua81}. In the present work we
neglect the three--nucleon potential.
For the $\la$--nucleon ($\la$N) potential we have chosen that
proposed by Bodmer and Usmani \cite{bod88}. 
We have investigated the effects of many-body forces involving the
$\la$ by explicitly considering a $\la N N$ interaction or, 
alternatively, by including a density dependence in the bare $\la N$
force. 

The interactions generate strong NN and $\la$N correlations, beyond 
those described by a mean field approach. In our work the correlation 
effects are accounted for by a Jastrow correlated wave function, 
where the two--body correlations depend on the interparticle distances 
only. More
realistic hamiltonians demand for a strong spin and isospin 
dependence in the correlation factor. However, 
we consider the simple Jastrow ansatz as 
a first step towards a more complete description of the hypernuclei. 
The same strategy has been followed for doubly closed shell nuclei, where 
the accuracy of the variational method has now reached the same level 
as in nuclear matter\cite{fab98}.

The FHNC equations for a $\la$ embedded in infinite nuclear
matter were developed in Ref.\cite{usm80}. In Ref.\cite{co92}, 
doubly closed shell nuclei in $ls$--coupling were studied by 
FHNC and Jastrow--type correlations, while the $jj$--coupling was 
introduced in such a scheme in Ref.\cite{ari96}. It was so 
possible to study nuclei ranging from $^{12}$C to $^{208}$Pb 
within the correlated basis function theory, 
using cluster expansion at all orders and avoiding such shortcuts as 
low order cluster truncations and local density approximations. 
The results obtained 
in those papers are the basis of the present FHNC study of 
$\Lambda$--hypernuclei. 

The plan of the paper is as follows: 
section \ref{hami} is devoted to a brief description of the hamiltonian 
and of the correlated wave function; section \ref{dens} gives an
outline of the FHNC theory for the hypernucleus; in section \ref{vari}
the expressions for the variational energy are presented and
discussed; the results are given in section \ref{resu} 
and the conclusions are drawn in section \ref{conc}.

\section{Hamiltonian and trial wavefunctions.}
\label{hami}
The system under study is a hypernucleus composed by $A$ nucleons and
one $\la$ hyperon. 
The formalism we have developed is however general enough to describe
any nuclear systems containing a non-nucleonic particle, 
treated as an impurity in the nucleonic fluid.

We write the hamiltonian of the system as: 
\begin{eqnarray}
H_{N+\Lambda}& = & H_N+H_{\Lambda}=\nonumber \\ 
& &  - \sum_{i=1}^A \frac{\hbar^2}{2m_i} \nabla^2_i
+\sum_{j>i=1}^A V^{NN}(i,j)  \nonumber \\ 
 & & - \frac{\hbar^2}{2m_\Lambda} \nabla^2_\Lambda
+\sum_{i=1}^A V^{\Lambda N}(\Lambda,i) +\sum_{j>i=1}^A 
V^{\Lambda NN}_3 (\Lambda,i,j)~, 
\label{H}
\end{eqnarray}
where we have considered two-- and three--body interactions
between the hyperon and the nucleons and only two--body ones between the 
nucleons. In the hamiltonian (\ref{H}) the purely nucleonic 
part, $H_N$, has been separated  
from that  involving the $\la$, $H_{\Lambda}$. 
The separation will be useful in the remainder of the paper.

As already stated in the introduction, we use the modified NN  S3 potential, 
which is fixed to reproduce the $s$ wave scattering
nucleon-nucleon phase shifts at low energies, gives reasonable 
results both for light nuclei and nuclear matter and allows 
for utilizing only central two--body correlations between the nucleons.
The operatorial structure of S3 is: 
\begin{equation}
\label{vn}
V^{NN}(i,j)= \sum_{p=1}^4 V^p(r_{ij}) O^p_{ij}~,
\end{equation}
where $O^{p=1-4}_{i,j}=1$, $\nsigma_i \cdot\nsigma_j$,
$\ntau_i \cdot \ntau_j$, $\nsigma_i \cdot \nsigma_j \ntau_i \cdot\ntau_j$. 
Modern nucleon-nucleon potentials are more complicated and contain the
essential tensor terms, absent in S3. On the other hand, 
we are mainly interested in presenting and testing the formalism. 
Adopting the S3 potential is sufficient for this purpose and has the 
advantage of avoiding the technical complications deriving from the use 
of more realistic potentials.

Within the same philosophy, we have chosen  the
$\la$--nucleon potential proposed by Bodmer and Usmani
\cite{bod88}. This potential has central, spin and space--exchange
components:  
\begin{equation}
\label{vl}
V^{\Lambda N}(1,2)=  V^1_{\Lambda N}(r_{12}) 
+ V^2_{\Lambda N}(r_{12}) \nsigma_{\Lambda} \cdot \nsigma_N  
+ V^x_{\Lambda N}(r_{12}) P_x~,
\end{equation}
with
\begin{eqnarray}
V^1_{\Lambda N}(r) &=& v_0(r) \left( 1 - \epsilon \right)~, \\
V^2_{\Lambda N}(r) &=& \frac {1} {4} (v_s-v_t) T_\pi^2 (r)~,  \\
V^x_{\Lambda N}(r) &=&  \epsilon ~ v_0(r)~,
\end{eqnarray}
where $P_x$ is the spatial exchange operator and $v_0$ is defined as:
\begin{equation}
v_0 (r) = \frac{W_c}{1+ \exp (\frac{r-R}{a})} - 
          \frac {1}{4} (v_s + 3v_t) T_\pi^2 (r)
\,\,\, .
\end{equation}
The one-pion exchange term is
\begin{equation}
T_\pi (r) = 
\left( 1 + \frac{3}{\mu r} + \frac {3}{(\mu r)^2} \right)
\frac{e^{-\mu r}}{\mu r} \left( 1 - e^{- c r^2} \right)^2
\,\,\, ,
\end{equation}
with the pion mass $\mu$=0.7 fm$^{-1}$ and the cutoff parameter  
$c$=2.0 fm$^{-2}$.
The values of the various parameters $v_s,v_t,\epsilon,W_c,R,a$ 
are given in Ref. \cite{usm95}. 
In this reference,
 a three-body $\la$-nucleon-nucleon interaction was added 
to the just described $\la$-nucleon one in order to fit the hypernucleus 
$^{17}_{~\Lambda}O$ empirical $B_\Lambda$ by the variational Monte Carlo (VMC) 
method. The $\la NN$ interaction 
has an important, attractive two-pion exchange part, which 
is the source of a tensor term. The tensor contribution, however, 
results to be zero if simple central correlations are 
adopted. As a consequence, such a semi-microscopic $\la NN$ interaction is 
too repulsive in a Jastrow correlated model, as we shall discuss later.  
For this reason, we simulate the effects of three-- and more--body 
potentials by using an approach similar to that developed by Friedman
and Pandharipande for pure nucleonic matter \cite{fri81}. 
We multiply $V^{\la N}(1,2)$ by a density
dependent function 
\begin{equation}
F_\rho(1,2)= \exp\{- \gamma [\rho(r_1)+\rho(r_2)]/2\}~, 
\label{Frho}
\end{equation}
where $\rho(r)$ is the nucleonic density. The value of the
parameter $\gamma$ is fixed to reproduce the empirical value of a
single $\la$ binding energy in nuclear matter, 
$B_\Lambda^{NM}=30$ MeV. 

The binding energy of the $\la$ particle is  
defined as minus the difference between the energies of the nuclear systems  
with and without the $\la$ :
\begin{eqnarray}
\label{blambda}
-B_\Lambda & = &  \frac 1 {2j+1} \sum_m <H_{N+\Lambda} >_{A+\Lambda} 
-<H_{N}>_{A} \nonumber \\
& & - \left(\frac 1 {2j+1} \sum_m <T^{cm}_{N+\Lambda} >_{A+\Lambda}
- <T^{cm}_{N} >_{A}\right)~, 
\end{eqnarray}
where we have used the notation:
\begin{equation}
< X >_Y = \frac 1 { <\Psi_Y | \Psi_Y >} \int d \ntau \Psi^*_Y X \Psi_Y~,
\end{equation}
with $Y=A+\la$ or $A$.
In the previous equations we have averaged on the third component of
the $\la$ total angular momentum and we have explicitly singled out 
the difference between the center of mass energies of the 
two systems. 

Our approach is based upon the variational principle, therefore the
energy functionals given in eq.(\ref{blambda}), 
$<H_{N+\Lambda} >_{A+\Lambda}$ and $<H_{N}>_{A}$,
 should be independently minimized with
respect to variations of the many-body wavefunctions, $\Psi_Y$. The
hypernucleus wavefunction is chosen as:
\begin{equation}
\label{psil}
|\Psi_{A+\Lambda}> = \left( \prod_{i=1}^A f_\Lambda (r _{\Lambda i}) \right)
\phi_{nljm}^\Lambda (\br_\Lambda)|\Psi_{A}>~,
\end{equation}
where $\phi_{nljm}^\Lambda$ is the mean-field single particle
wavefunction of the $\la$ in $jj$--coupling, 
$f_\Lambda (r _{\Lambda i})$ is a two-body scalar (Jastrow)  
correlation between the hyperon and a single nucleon and $|\Psi_{A}>$ is 
the correlated wave function describing the remaining $A$ nucleons. In
our calculation this function is defined as:
\begin{equation}
\label{psin}
|\Psi_{A}> = \left( \prod_{j>i=1}^A f_N(r_{ij}) \right)
 \Phi_{N,Z}(1,\ldots,A)~,
\end{equation} 
where $\Phi_{N,Z}$ is the Slater determinant of a set of 
single particle wave functions of $N$ neutrons and $Z$ protons. 
Obviously, $A=Z+N$.

\section{Densities and FHNC equations}
\label{dens}
The energy expectation values are evaluated by means of the 
Fermi hypernetted chain\cite{ros82} cluster diagrams 
resummation technique. As already mentioned, our approach
consists in considering the $\la$ as an impurity in the nucleonic
fluid. The FHNC equations for an impurity in homogeneous matter 
were derived in Ref. \cite{usm80} for a $\Lambda$ hyperon in 
symmetric nuclear matter and in Refs. \cite{fab82,bor94} 
for a $^4$He atomic impurity in liquid $^3$He. 

Our aim is the evaluation the $\la$ binding energy in finite nuclear
systems. 
To this purpose, we extend the FHNC equations developed in 
Ref.\cite{ari96} for doubly closed shell nuclei
in the $jj$--coupling scheme with a  Jastrow correlated wave 
functions. 

\begin{figure}
\begin{center}
\vspace*{1.cm}
\hspace*{-2.0 cm}
\leavevmode
\epsfysize = 350pt
\epsfbox[70 200 500 650]{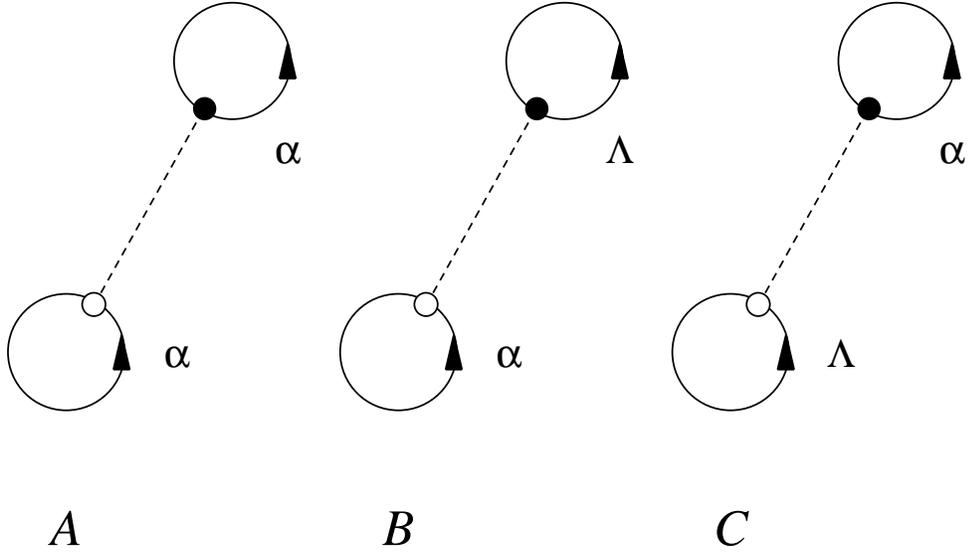}
\end{center}
\vskip 3.5 cm
\caption{  
  Examples of cluster diagrams contributing to the one body
  densities. The A diagram contributes to $\rho_{1,A}^\alpha (r)$, the
  B diagram to $\rho_{1,\Lambda}^\alpha (r)$ and the C diagram to 
  $\rho_1^{\Lambda} (r)$. The dashed lines represent 
  the dynamical correlations, 
  $f^2_D(r_{ij})-1$, ($D=N$ in A and $D=\Lambda$ the rest of the cases) 
  and the oriented lines the statistical correlations, 
  $\rho_0(\br_i, \br_j)$. A black dot associated with a point implies 
  integration over its coordinates. The label $\alpha$ indicates the
  nucleon. 
}
\label{fig:diag}
\end{figure}

The basic quantities to be analyzed are the one- and two-body densities 
(OBD and TBD). It is convenient to define densities for each type of 
particle, protons $p$, neutrons $n$ and $\la$:
\begin{eqnarray}
\label{ro1}
\rho_{1,A+\Lambda}^\alpha (\br ) & = &  \langle
\sum_{k=1}^A \delta (\br - \br _k) P^{\alpha}_k \rangle_{A+\Lambda} \\
\label{ro1r}
\rho_1^{\Lambda} (\br ) & = & \langle
\delta (\br - \br _{\Lambda}) \rangle_{A+\Lambda} 
\end{eqnarray}
where $\alpha=p,n$, $P^\alpha$ is the projector operator over the
particle of $\alpha$ type. An averaged sum on the third
components of the $\la$ particle angular momentum  is
understood.  
As in Refs.\cite{co92,ari96} the densities are evaluated by using
cluster expansion techniques.
The nucleonic OBD are divided in two parts:
\begin{equation}
\label{ro1l}
\rho_{1,A+\Lambda}^\alpha (\br ) 
= \rho_{1,A}^\alpha (\br ) + \rho_{1,\Lambda}^\alpha (\br ).
\end{equation}
The first part is given by the sum of the  cluster diagrams containing  
only nucleons, and represents the nuclear bulk contribution 
to the OBD. One of the diagrams contributing to this part is 
the $A$ diagram of Fig.\ref{fig:diag}. 
The second part of eq. \ref{ro1l}, called rearrangement term, is
obtained by summing all the diagrams containing the $\la$ as an internal
particle, like the diagram $B$ of Fig.\ref{fig:diag}, and provides 
the modification of the nucleon OBD due to the presence of the 
$\Lambda$ impurity. A mean-field description of the hypernucleus does 
not provide any rearrangement term.  Finally, 
the diagrams where the $\la$ is an external particle, like the $C$
diagram of Fig.\ref{fig:diag}, are summed in the \la-OBD, 
$\rho_1^{\Lambda} (\br)$. 
The separation of the one-body densities in bulk, rearrangement
and $\la$ terms will be very useful for the calculation of the
$\la$ binding energy in the nucleus and it will be exploited also for 
the calculation of other quantities occurring in the FHNC scheme. 

In analogy with the one--body densities, we define the two--body
densities as:
\begin{eqnarray}
\label{ro2n}
\rho_{2,q,A+\Lambda}^{\alpha \beta}(\br ,\br') & = & 
\langle 
\sum_{k \ne l=1}^A \delta (\br  - \br _k)
\delta (\br' - \br _l)O^{q}_{kl}P^{\alpha}_k P^{\beta}_l
\rangle_{A+\Lambda} \\
\label{ro2r}
& = & \rho_{2,q,A}^{\alpha \beta}(\br ,\br') +
\rho_{2,q,\Lambda}^{\alpha \beta}(\br ,\br')~, \nonumber
\end{eqnarray}
where $\alpha,\beta=p,n$ and
$O^{q}_{kl}$ is one of the operators defined in eq.(\ref{vn}).
The $\la$N two--body densities  are defined as:
\begin{eqnarray}
\label{ro2l}
\rho_{2,q}^{\Lambda \alpha}(\br ,\br') & = & 
\langle 
\delta (\br  - \br _{\Lambda}) \sum_{k=1}^A 
\delta (\br' - \br _k)O^{q}_{\Lambda k}P^{\alpha}_k 
\rangle_{A+\Lambda}~, \\
\label{ro2x}
\rho_{2,x}^{\Lambda \alpha}(\br ,\br') & = & 
\langle 
\delta (\br  - \br_{\Lambda}) \sum_{k=1}^A 
\delta (\br' - \br_k) P_x P^{\alpha}_k 
\rangle_{A+\Lambda}~, 
\end{eqnarray}
where $O^{q}_{\la k}=1,~\nsigma_{\Lambda} \cdot \nsigma_k$ are 
the operators of the $\la$N interaction (\ref{vl}),  
and $P_x$ is the space exchange operator between the $\la$ and the
k-nucleon.  

In the FHNC approach the TBD are described in
terms of two-body correlation functions 
$f_N(r_{kl})$ and $f_{\Lambda} (r_{\Lambda k})$, and of uncorrelated
densities. The latter are built from a single particle basis, defined 
in the present paper as a set of $jj$ coupled wave functions of the type:
\begin{equation}
\label{phi}
\phi_k^t(x) \equiv
\phi^{t}_{nljm}(\br)= R^{t}_{nlj}(r)\sum_{\mu,s} 
<l \mu \half s | j m> Y_{l,\mu}(\what{r}) \chi_s~, 
\end{equation}
where $t=p,n,\la$.  $Y_{l,\mu}$ are the spherical harmonics,
$<l \mu \half s | j m>$ are the Clebsch--Gordan coefficients and $\chi_s$
the two-component Pauli spinors. 

In the nucleonic case, the uncorrelated two--body densities are:
\begin{eqnarray}
\nonumber
\rho_0^\alpha (\br_1, \br_2) 
&=& \sum_k \phi_k^{\alpha *} (\br_1) \phi_k^\alpha (\br_2) \\
\label{ro20n}
&=& \sum_{s,s'} \rho^{s,s',\alpha}_0(\br_1,\br_2)
\chi_s^\dagger (1)  \chi_{s'} (2)~.
\end{eqnarray}
The explicit expressions of the 
parallel, $s=s'$, and antiparallel $s=-s'$ 
densities are given in Ref. \cite{ari96} together with  
the other purely nucleonic FHNC quantities.

We briefly discuss here the quantities involving the $\la$.
The uncorrelated $\la$ one--body density is given by:
\begin{equation}
\rho^{\Lambda}_0(\br) =   \frac 1 {2j+1} \sum_m \phi^{\Lambda *}_{nljm}(\br)
\phi^{\Lambda}_{nljm}(\br) = \frac{1} {4 \pi} 
\left( R^{\Lambda}_{nlj}(r) \right)^2~.
\end{equation}

The one-body $\la$ uncorrelated density matrix is needed to calculate
diagrams where the spatial coordinates between the $\Lambda$ and one
nucleon are exchanged. Its expression is:
\begin{equation}
\rho^{\Lambda}_0(\br_1, \br_2) = \frac{1} {4 \pi} 
R^{\Lambda}_{nlj}(r_1) R^{\Lambda}_{nlj}(r_2) P_l(\cos \theta_{12})~,
\end{equation}
where we have indicated 
with $P_l$ the Legendre polynomial of $l$th degree and $\theta_{12}$ the
angle between the two vectors. 

A detailed description of the FHNC equations in the nucleonic
matter, in a $jj$ coupled single particle basis computational scheme, 
has been done in  Ref. \cite{ari96}. We do not give here the
equations for this case, but we rather present and discuss the changes
related to the presence of the $\la$. The FHNC \la--densities are:
 \begin{eqnarray}
 \rho_1^{\Lambda} (r) & = & \xi_d^{\Lambda} (r) = 
 \xi_e^{\Lambda} (r) \rho_0^{\Lambda} (r)~, \\
 \xi_e^{\Lambda} (r) & = & \frac 1 {C_{\Lambda}} \exp [U_d^{\Lambda} (r)]~, \\ 
 C_{\Lambda} & = & \int d \br~\rho_0^{\Lambda} (r)~\exp [U_d^{\Lambda} (r)]~, \\
 \rho_{2,1}^{\Lambda \beta} (\br_1, \br_2) & = &  \xi_d^{\Lambda} (r_1) \left(
 \xi_d^\beta (r_2) g_{dd}^{\Lambda \beta} (\br_1, \br_2) +
 \xi_e^\beta (r_2) g_{de}^{\Lambda \beta} (\br_1, \br_2) \right)~, \\
 \label{r=0}
 \rho_{2,2}^{\Lambda \beta} (\br_1, \br_2) & = & 0 ~.
 \end{eqnarray}

The functions $\xi_{d,e}^\Lambda(r)$, $U_{d}^\Lambda(r)$ and 
$g^{\Lambda \beta}_{xy} (\br_1, \br_2)$ are analogous to their
nucleonic partners whose expressions are given  in the Appendices A
and B of Ref. \cite{ari96}, and they are obtained   
with the substitution $\alpha\rightarrow\Lambda$. 
In extending the expressions given in the reference, we should
remember that there is no sum on $\la$, 
since it is just a single external impurity, 
and that we consider the same correlation function between
$\la$ and any type of nucleon: 
$f_{\Lambda \beta} (r_{12})=f_{\Lambda} (r_{12}) $ with $\beta=p,n$.

Eq.(\ref{r=0}) shows that, in our case, the spin component of the 
two-body density, $\rho_{2,2}^{\Lambda \beta}$, is zero. 
The reason lies in the fact that only diagrams corresponding to 
exchanges between identical particles contribute to the 
spin component of the two--body density for Jastrow correlated wave
functions. As a consequence they do not affect the $\la$N--TBD. 
Direct diagrams, where the particles are not exchanged,
may contribute to $\rho_{2,2}^{\Lambda \beta}$ only if the 
correlation contains spin dependent components (which is not our case). 
Moreover, this type of contribution has been found to be generally 
small \cite{usm95}.

The evaluation of the diagrams containing the
space exchange $\la$N potential requires 
the introduction of a space exchange density, given by:
\begin{equation}
\rho_{2,x}^{\Lambda \alpha}(\br ,\br') = \frac{2}{C_{\Lambda}} 
f_{\Lambda}^2 (|\br - \br'|) 
\rho^{\Lambda}_0(\br, \br') \rho^{s s \alpha}(\br , \br')~,
\end{equation}
where $\rho^{s s \alpha}(\br , \br')$ is defined as
in Appendix B of \cite{ari96}, with the substitutions 
$f_{\alpha \beta}(r_{12})\rightarrow f_{\Lambda} (r_{12})f_N(r_{12})$ 
in the construction of the $g_{\omega d}$ functions.

The expressions of the rearrangement terms are quite lengthy since the
$\la$ may be present in any of the elements involved in the densities.
We give below the expression of the contributions to the one--body density:
\begin{eqnarray}
\rho_{1,\Lambda}^\alpha (r) & = & 
\left( U_{d,\Lambda}^\alpha (r) 
\left[ \rho_0^\alpha (r) + U_e^\alpha (r) \right] +
U_{e,\Lambda}^\alpha (r) \right)~\exp [U_d^\alpha (r)] \\
 & = & \xi_d^\alpha (r) U_{d,\Lambda}^\alpha (r) +\xi_e^\alpha (r) 
U_{e,\Lambda}^\alpha (r) = \xi_{d,\Lambda}^\alpha (r)~, \\
\xi_{e,\Lambda}^\alpha (r) & = &  
U_{d,\Lambda}^\alpha (r)~\exp [U_d^\alpha (r)]~. 
\end{eqnarray}
The $U_{X=d,e}^\alpha$ functions are given in Ref.\cite{ari96}, 
while the $U_{X,\Lambda}^\alpha$ and the 
rearrangement parts of the two--body densities are given in the
Appendix. 

\section{The variational energy}
\label{vari}
In this section we discuss the evaluation of the variational energy of 
a $\la$ hypernucleus in the framework of the FHNC approach. 
The potential energy is separated into its nucleonic and $\la$ pieces:
\begin{equation}
<V_{N+\Lambda} >_{A+\Lambda} = <V_{N} >_{A+\Lambda} + 
<V_{\Lambda}>_{A+\Lambda} ~.
\end{equation}
The expectation values can be expressed in terms of the two--body distribution 
functions as:
\begin{eqnarray}
 <\overline V_{N}>_{A+\Lambda}  & = & \half 
\sum_{\alpha,\beta=p,n} \sum_{q=1}^4 \int 
d \br_1 d \br_2 V^q (r_{12}) \rho_{2,q,A+\Lambda}^{\alpha \beta}(\br_1
,\br_2) \nonumber \\
= \ \ \ <V_{N} >_{A} & + & \half \sum_{\alpha,\beta=p,n} \sum_{q=1}^4 \int 
d \br_1 d \br_2 V^q (r_{12}) \rho_{2,q,\Lambda}^{\alpha \beta}(\br_1
,\br_2)~~, \\
<\overline V_{\Lambda}>_{A+\Lambda} & = & 
\sum_{\alpha=p,n} 
\sum_{q=1}^2 \int d \br_1 d \br_2 V^q_{\Lambda N} (r_{12}) 
\rho_{2,q}^{\Lambda \alpha}(\br_1,\br_2) + \nonumber \\
&  & \sum_{\alpha=p,n} \int d \br_1 d \br_2 V^x_{\Lambda N} (r_{12}) 
\rho_{2,x}^{\Lambda \alpha}(\br_1,\br_2)~, 
\end{eqnarray}
where $<\overline { V}>$ indicates the
average on the third components of the $\la$ angular momentum and 
$<V_{N} >_{A} $ is the bulk nucleonic potential energy.

Because the minimizations of $<H_{N+\Lambda} >_{A+\Lambda}$ and 
$<H_{N} >_{A}$ must be, in principle, carried on independently, 
the nucleon--nucleon Jastrow correlation, $f_N(r)$, might be different in 
the two cases, as well as the nucleon single particle wave functions, 
$\phi_k({\bf r})$. However, we found that in our cases $f_N(r)$ and 
$\phi_k({\bf r})$ do not practically change in going from the $A$ to 
the $A+\Lambda$ system. As a consequence, in our approach the nucleonic 
part of the hypernucleus wave function has been kept the same as in nucleus. 
This fact allows for obtaining the potential energy contribution 
to the binding energy of the $\Lambda$ by subtracting the pure nucleus 
$<V_{N} >_{A}$ from $<\overline V_{N}>_{A+\Lambda}$. The remaining part 
is separated in  two contributions: the potential energy due to 
the interaction  $\Lambda$--nucleon, or interaction energy, 
and the modifications of the nucleon--nucleon potential energy due 
to the presence of the $\Lambda$, or rearrangement energy. 
We put in evidence these contributions by rewriting:
\begin{equation}
V_{\Lambda} =  
<\overline { V}_{N+\Lambda}> _{A+\Lambda}-<V_{N} >_{A} = 
V_{\Lambda}^I + V_{\Lambda}^R~,
\end{equation}
with
\begin{equation}
V_{\Lambda}^I  =   \sum_{\alpha=p,n} 
\int d \br_1 d \br_2 \left(\sum_{q=1}^2 V^q_{\Lambda N} (r_{12}) 
\rho_{2,q}^{\Lambda \alpha}(\br_1,\br_2) + V^x_{\Lambda N} (r_{12}) 
\rho_{2,x}^{\Lambda \alpha}(\br_1,\br_2) \right)~, \\
\end{equation}
and
\begin{equation}
V_{\Lambda}^R =  \half \sum_{\alpha,\beta=p,n} \sum_{q=1}^4 \int 
d \br_1 d \br_2 V^q (r_{12}) \rho_{2,q,\Lambda}^{\alpha \beta}(\br_1
,\br_2)~. 
\end{equation}

The separation in interaction and rearrangement terms can also be done
for the $\la$ kinetic energy, evaluated here  by means of the 
Jackson--Feenberg expression \cite{co92}. Following an analysis similar 
to that done for the potential energy, we obtain:
\begin{equation}
T_{\Lambda} = 
<\overline { T}_{N+\Lambda} >_{A+\Lambda}-<T_{N} >_{A} = 
T_{\Lambda}^I + T_{\Lambda}^R~.
\end{equation}
The interaction kinetic energy is given by:
\begin{eqnarray}
T_{\Lambda}^I & = &  -\frac{\hbar^2}{4m_{\Lambda}}
\int d \br d \rho_{T1}^{\Lambda}(\br) \xi_e^{\Lambda} (\br) \nonumber \\
& & -\frac{\hbar^2}{4} \left( \frac{1}{m}+\frac{1}{m_{\Lambda}} \right)
 \sum_{\alpha=p,n}  \int d \br_1 d \br_2 ~t[f_{\Lambda} (r_{12})] 
~\rho_{2,1}^{\Lambda \alpha}(\br_1,\br_2) \,\,\, ,
\end{eqnarray}
with 
\begin{equation}
\rho^{\Lambda}_{T1}(r_1) = \frac{1} {4\pi}
\left[ R^{\Lambda}_{nlj}(r_1)\left( D^{\Lambda}_{nlj}(r_1) - 
\frac{l(l+1)}{r^2_1} R^{\Lambda}_{nlj}(r_1) \right)
- \left( R^{\Lambda \prime}_{nlj}(r_1) \right)^2 \right] \,\,\, .
\end{equation}
The expressions of $t[f_{\Lambda}]$ and 
$D^\alpha_{nlj}$ can be obtained from Ref.\cite{ari96} (Eq.(16) and
Appendix C).

The rearrangement part of the kinetic energy is:
\begin{eqnarray}
T_{\Lambda}^R & = & - \frac{\hbar^2}{4m}
 \sum_{\alpha,\beta=p,n}  \int d \br_1 d \br_2 t[f_N (r_{12})] 
\rho_{2,1,\Lambda}^{\alpha \beta}(\br_1,\br_2) \nonumber \\
&  & -\frac{\hbar^2}{4m} \sum_{\alpha=p,n}  \int d \br_1 
\rho_{T1}^{\alpha}(\br_1) \xi_{e,\Lambda}^\alpha (\br_1)  \\
& & +\frac{\hbar^2}{4m}
 \sum_{\alpha=p,n}  \int d \br_1 d \br_2
\rho_{T2}^{\alpha}(\br_1, \br_2) \left[
\xi_e^\alpha (\br_1) \xi_e^\alpha(\br_2) 
g_{dd,\Lambda}^{\alpha \alpha} (\br_1,\br_2) + \right. \nonumber \\ 
& &  
\left. \left(\xi_{e,\Lambda}^\alpha (\br_1) \xi_e^\alpha(\br_2)+ 
\xi_e^\alpha (\br_1) \xi_{e,\Lambda}^\alpha(\br_2) \right) 
g_{dd}^{\alpha \alpha} (\br_1,\br_2) \right] \nonumber \\
& & - \frac{\hbar^2}{2m}
\sum_{\alpha=p,n}  \int d \br_1 d \br_2 \Bigl[
\rho_{T3,P}^{\alpha}(\br_1, \br_2) H_{cc,P,\Lambda}^{\alpha}(\br_1, \br_2)
\nonumber  \\
& & 
+\rho_{T3,A}^{\alpha}(\br_1, \br_2) 
H_{cc,A,\Lambda}^{\alpha}(\br_1, \br_2) \Bigr]~,   \nonumber 
\end{eqnarray}
where 
\begin{eqnarray}
H_{cc,D,\Lambda}^\alpha (\br_1, \br_2)) & = & \xi_{e,\Lambda}^\alpha (r_1) \Bigl[
\xi_e^\alpha (r_2) \left( \left( g_{dd}^{\alpha \alpha} (\br_1, \br_2) -1 \right)
N_{cc,D}^\alpha (\br_1, \br_2)) + N_{cc,D}^{(x) \alpha} (\br_1, \br_2) \right)  
\nonumber \\
& + & \left(\xi_e^\alpha (r_2) -1 \right)
 N_{cc,D}^{(\rho) \alpha} (\br_1, \br_2)+ 
\xi_e^\alpha (r_2)  g_{dd}^{\alpha \alpha} (\br_1, \br_2) 
E_{cc,D}^\alpha (\br_1, \br_2)  \Bigr] \nonumber \\
& + &  \xi_e^\alpha (r_1)\xi_{e,\Lambda}^\alpha (r_2) 
g_{dd}^{\alpha \alpha} (\br_1, \br_2) 
\left( N_{cc,D}^\alpha (\br_1, \br_2) +E_{cc,D}^\alpha (\br_1, \br_2) \right) 
\nonumber \\
& + &  \xi_{e}^\alpha (r_1) \Bigl[
\xi_e^\alpha (r_2) \Bigl( \left( 
g_{dd}^{\alpha \alpha} (\br_1, \br_2) -1 \right) 
N_{cc,D,\Lambda}^\alpha (\br_1, \br_2)  
 \\
& + & g_{dd,\Lambda}^{\alpha \alpha} (\br_1, \br_2)  
N_{cc,D}^\alpha (\br_1, \br_2) +
 N_{cc,D,\Lambda}^{(x) \alpha} (\br_1, \br_2) \Bigr) \nonumber \\
& + & \left(\xi_e^\alpha (r_2) -1 \right)  
N_{cc,D,\Lambda}^{(\rho) \alpha} (\br_1, \br_2)
\nonumber \\ 
& + &  \xi_e^\alpha (r_2) \left(  
g_{dd,\Lambda}^{\alpha \alpha} (\br_1, \br_2) E_{cc,D}^\alpha (\br_1, \br_2) +
g_{dd}^{\alpha \alpha} (\br_1, \br_2) E_{cc,D,\Lambda}^\alpha (\br_1, \br_2) 
\right) \Bigr] \nonumber .
\end{eqnarray}
Again the expressions of $\rho_{Tk}^\alpha$ for $k=1,2,3$
can be found in the appendix C of \cite{ari96}.

The difference of the center of mass kinetic energies is given by:
\begin{equation}
T_{\Lambda}^{cm} =  -\frac{m_{\Lambda}}{Am+m_{\Lambda}} <T_{N}^{cm}>_A 
-\frac{\hbar^2}{4(Am+m_{\Lambda})}\int d \br_1  \rho_{T1}^{\Lambda}(\br_1)~, 
\end{equation}
and the expression of $<T_{N}^{cm}>_A$ is given in eq. (23) of 
Ref.\cite{ari96}.

\section{Results}
\label{resu}
We have studied the ground state structure of those \la--hypernuclei having 
$N$ and $Z$ values corresponding to doubly closed shell nuclei in 
$jj$--coupling, $^{13}_{~\Lambda}$C, $^{17}_{~\Lambda}$O, 
$^{41}_{~\Lambda}$Ca, $^{49}_{~\Lambda}$Ca,
$^{91}_{~\Lambda}$Zr and $^{209}_{~~\Lambda}$Pb.  
The same set of nuclei where the neutron in the highest neutron single
particle level has been substituted by a $\la$ hyperon
($^{12}_{~\Lambda}$C, 
$^{16}_{~\Lambda}$O, $^{40}_{~\Lambda}$Ca, 
$^{48}_{~\Lambda}$Ca, $^{90}_{~\Lambda}$Zr and $^{208}_{~~\Lambda}$Pb)  
have been also considered. 
In order to deal in FHNC with the partially occupied neutronic level, 
we adopt the 
following procedure: the degeneracy factor $2j+1$, multiplying 
the partially occupied state contribution to the uncorrelated nucleonic OBD, 
has been replaced by the factor $2j$. In this way, the OBD results 
normalized to $A-1$.
Moreover, in the spin parallel part of the uncorrelated nucleonic TBD,
$\rho^P$, the $2j+1$ factor has been substituted by 
$2 \sqrt{(2l+1)(2l+1/2)}-4l+2j-1$, while the (much smaller) antiparallel 
part, $\rho^A$, remains unchanged. 
This choice ensures a correct normalization of the TBD. 

Two different mean field potentials have been used to build the
single--particle nuclear wave functions. The first potential is a 
harmonic oscillator (HO) well with the same constant, $b_N$, for
protons and neutrons; the second choice is the Woods--Saxon (WS)
potential used in the calculations of Ref.\cite{ari96}. The parameters 
of the WS potential are different for protons and neutrons.
We have always used a harmonic oscillator potential 
for the $\Lambda$ single--particle potential. 

The correlation functions have a gaussian form:
\begin{equation} 
f_{X=\Lambda ,N} (r) = 1 - \alpha_X \exp ( -\beta_X r^2 )~.
\label{corr}
\end{equation}

In order to calculate $B_\Lambda$, we first minimize the bulk 
energy of the nucleus, $<H_N>_A$.  
The minimization has been performed for each 
nucleus in two ways, depending on the nucleonic single--particle used. 
In the case of the HO potential, the nuclear correlation, 
$f_N(r)$, has been taken from nuclear matter and the nucleus energy has 
been obtained 
by varying only the values of the oscillator constant of the HO mean field. 
For the Woods--Saxon case, the mean field potential has been kept fixed 
as in Ref. \cite{ari96} and the parameters of the correlation 
have been varied.  
We consider the WS model as the most realistic one, since its parameters 
have been determined to reproduce at best the nuclear densities. 
It must be noticed that, in the region of the variational 
space corresponding to the energy minimum, the energy itself is rather 
insensitive to small changes of the WS parameters. 
Therefore it is possible to reproduce the densities and the radii
without excessively spoiling the quality of the minimum. 
 
For the description of the nucleonic part of
the hypernucleus we take the same $f_N(r)$ and mean field potential as 
in the corresponding nucleus. Then,
 $<H_{N+\Lambda}>_{A+\Lambda}$ is minimized 
by changing only the $\Lambda$--HO constant and the parameters of the 
$\Lambda$--nucleon correlation function, $f_\Lambda (r)$. 

The parameters of the nuclear correlation (\ref{corr}) with the HO 
mean field are $\alpha_N=0.7$ and $\beta_N=2.0$ fm$^{-2}$. 
The nuclear matter binding energy per nucleon with this correlation and 
the S3 potential, at saturation density ($\rho_{NM}=0.16$ fm$^{-3}$), 
is $B_{NM}$=14.43 MeV, close enough to the empirical value, 
$B_{NM,{\rm emp}}$=16 MeV,  
and comparable with the best, more sophisticated potentials on the 
market.  For each nucleus the value of the oscillator constant,
 $b_N$, has been fixed to get the energy minimum.
In Table \ref{tab:oscc} we compare our HO and WS binding energies and
rms charge radii with their experimental values.
A general reasonable agreement is
found, considering the relative simplicity 
of both wave functions and interactions. 
In particular, the WS radii are close the experimental ones.  The
$^{12}$C nucleus represents somehow an anomaly, having a good estimate
of the radius but a very small binding energy.  
The origin of this disagreement is still under investigation. It may 
be related to the inadequacy of a spherical model description of this
nucleus.

The parameter $\gamma$ of the 
density dependent potential (DDP), described in Section II, has been 
chosen to reproduce the $\Lambda$ binding energy in nuclear matter, 
$B_{\Lambda,NM}$. We find  $\gamma=2.2$ fm$^3$ and 
$B_{\Lambda,NM} = 30.26$ MeV and this value of the parameter $\gamma$ 
has been used in the remaining finite systems calculations.  
The parameters of the $\Lambda$--HO mean fields and of the $f_\Lambda(r)$ 
correlations at the $<H_{N+\Lambda}>_{A+\Lambda}$ minimum are shown 
in Table \ref{tab:osck}.

In Table \ref{tab:detail} we compare the contributions to the 
 $\Lambda$ binding energy in the doubly closed shell hypernuclei 
calculated with the HO nucleonic mean field with those evaluated 
in nuclear matter 
 We have also separated the interaction (I) and rearrangement (R) 
terms in the kinetic and potential energies.  The upper part of the 
table gives the results with the two--body $\Lambda$N interaction only 
and the related $\la$ binding energy, $B_\Lambda^{(2)}$. 
In the lower part of the table the contribution of the three-body 
$\la$NN interaction of Ref.\cite{usm95} and the corresponding binding
 energy  $B_\Lambda^{(3)} $ are shown. Finally we present the results
 obtained with the DDP potential.

Since the empirical value of  $B_\Lambda$ in $^{17}_{~\Lambda}$O
is estimated to be about 13 MeV \cite{usm95},  the 
result obtained with the $\la$N force only is too attractive. The
VMC  calculation of Ref. \cite{usm95} gives 
$B_{\Lambda,{\rm VMC}} ^{(2)} (^{17}_{~\Lambda} {\rm O})$=27.5$\pm$2.0
MeV,  
in qualitative agreement with our value. 
The empirical binding energy was reproduced in the VMC approach  
by the inclusion of an explicit $\la$NN force. Using the same three--body 
potential, we obtain 
$B_{\Lambda} ^{(3)} (^{17}_{~\Lambda} {\rm O})$=6.81 MeV. 
We have already pointed out that central, Jastrow, correlations
underestimate the  
attractive contribution of the $\Lambda$NN interaction, as clearly appears 
from the $B^{(3)}_\Lambda$ value. This potential, when used in
conjunction with  
Jastrow correlated wave functions, does not even bind the $\Lambda$ in
heavy  
nuclei. Tensor--like correlations are needed in order 
to effectively use potentials induced by one-- and more--pion exchanges in 
spherically symmetric systems\cite{fab98}. 
The introduction of a DDP, fitted to the 
$\Lambda$ binding in nuclear matter, brings $B_\Lambda$ reasonably 
close to the empirical estimate, even with a relatively simple 
correlation. All the results presented hereafter have been obtained 
with the DDP.

In Table \ref{tab:sho} we give the binding energies of a $\Lambda$ in its 
$1s$ ground state, calculated
with the HO single particle potential for several 
hypernuclei and  nuclear matter. We explicitly show the overall 
interaction and rearrangement contributions.  
The difference of the center of mass kinetic energies has been included 
in the rearrangement part. 
The increase of the $\la$ binding energy along A is mostly produced by the 
interaction energy and, specifically, by its potential energy part.
In contrast, the dependence of the rearrangement energy on $A$ is much
weaker. We found an analogous behavior the $\Lambda$ energies in the
$1p$ and $1d$ states.
To complete the information we give in Table \ref{tab:pho}, 
the results for the $\Lambda$ energies in the $1p$ and $1d$ states
with the nucleonic HO mean field.
In Table \ref{tab:ws} we compare
the $\la$ binding energies obtained with a Woods--Saxon nucleonic mean
field with the experimental energies. For the $^{90}_{~\Lambda}$Zr the
comparison is done with the energies measured in $^{89}_{~\Lambda}$Y.

\begin{figure}
\begin{center}
\vspace*{1.cm}
\hspace*{-2.0 cm}
\leavevmode
\epsfysize = 350pt
\epsfbox[70 200 500 650]{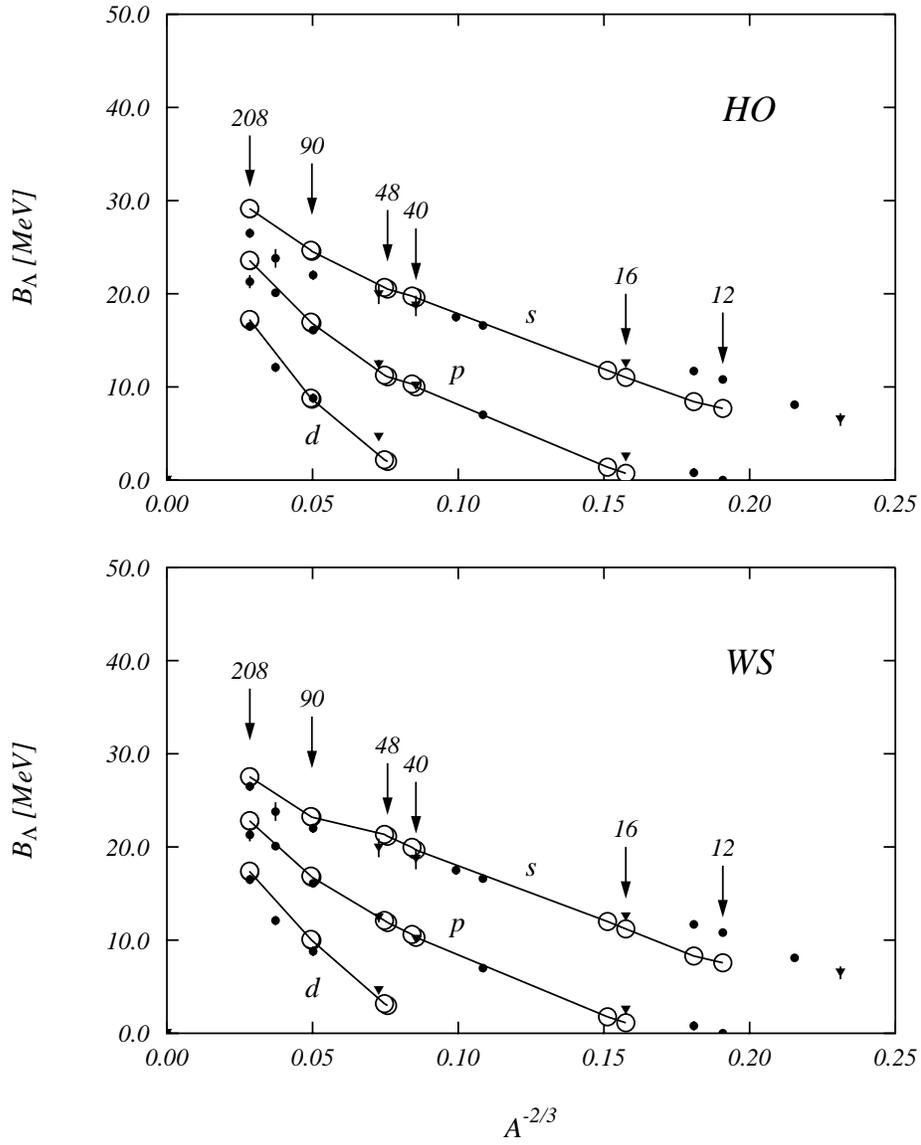}
\end{center}
\vskip 3.5 cm
\caption{ $\Lambda$ binding energies for the $1s$, $1p$ and $1d$ states,
as function of $A^{-2/3}$. The white circles are the energies
calculated using harmonic oscillator (upper panel) and Woods-Saxon
mean field potentials (lower panel). 
The experimental energies are taken from Ref. 
\protect\cite{has96} (dots) and 
\protect\cite{pil91} (triangles). 
The full lines connecting the theoretical values have been drawn to
guide the eyes.} 
\label{fig:ene}
\end{figure}

The results for the $\la$ binding energies are summarized in
Fig.\ref{fig:ene}, where they are presented as a function of $A^{-2/3}$
and compared with the experimental
energies of Refs. \cite{has96} (dots), \cite{pil91} (triangles)
and \cite{may97} for the $^{13}$C. 
The figure gives the $\la$ energies in the $1s$, $1p$ and $1d$ states. 
The quality of the agreement with the experiment
is rather good, in spite of the simplicity of the model. 
By inspecting the figure in more details, we find that all the calculations
underbind the $\la$ in Carbon by $\sim 30\%$, 
and provide a steeper variation of $B_\Lambda$ with respect to the 
experiments. This behavior resembles that already 
described in Carbon nucleus. 
It is also worth noticing that, 
as expected, the heavier nuclei, Zr and Pb, are better
described by the Woods-Saxon well than by the HO one.

\begin{figure}
\begin{center}
\vspace*{1.cm}
\hspace*{-2.0 cm}
\leavevmode
\epsfysize = 350pt
\epsfbox[70 200 500 650]{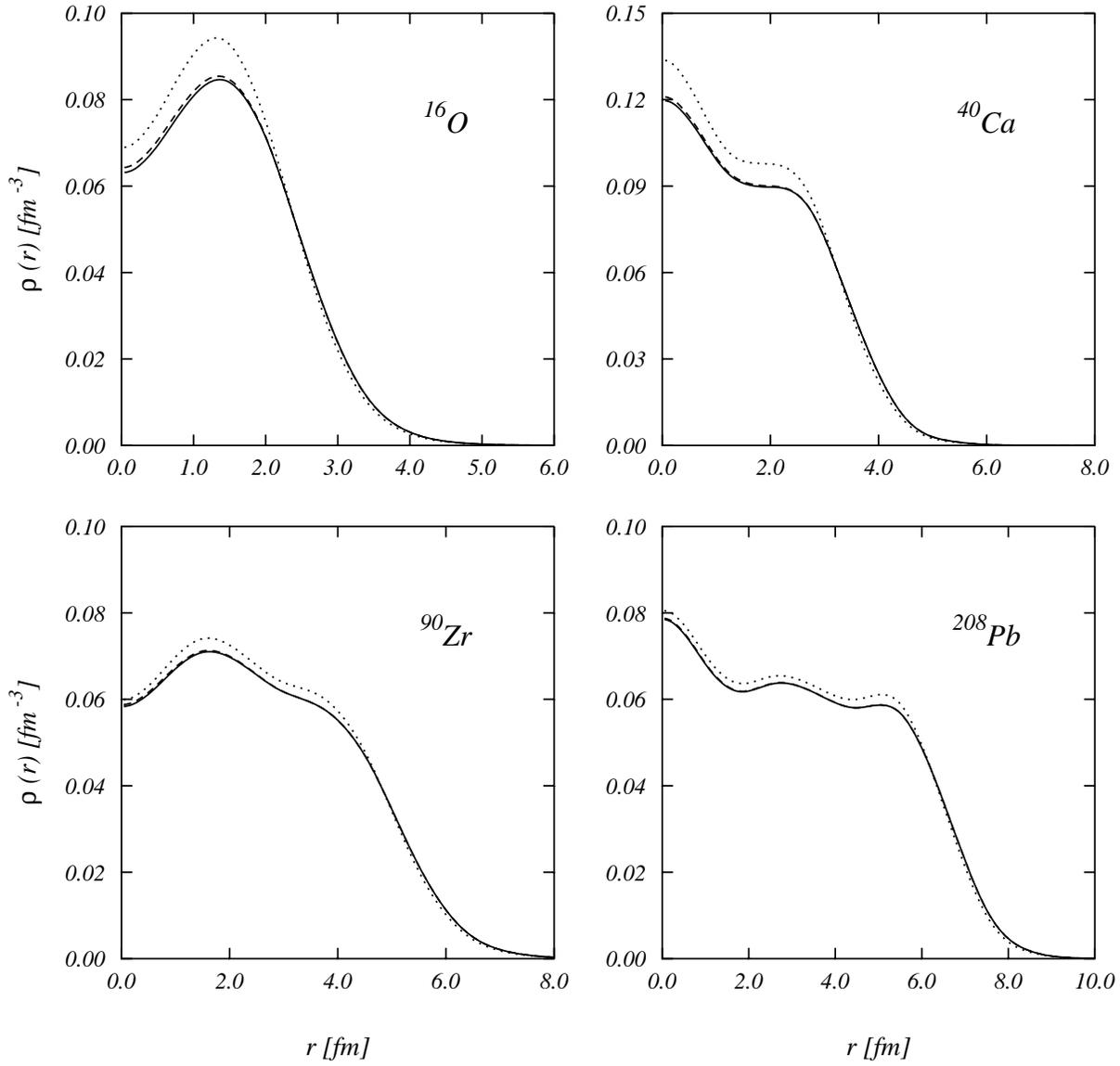}
\end{center}
\vskip 3.5 cm
\caption{ Proton densities for four doubly closed shell nuclei. 
  The dotted lines are the IPM results; the dashed
  lines are the densities obtained by purely nucleonic FHNC calculations;
  the full lines are the densities obtained when a $\la$ in the $s$ state is 
  added.}
\label{fig:dnuc}
\end{figure}

\begin{figure}
\begin{center}
\vspace*{1.cm}
\hspace*{-2.0 cm}
\leavevmode
\epsfysize = 350pt
\epsfbox[70 200 500 650]{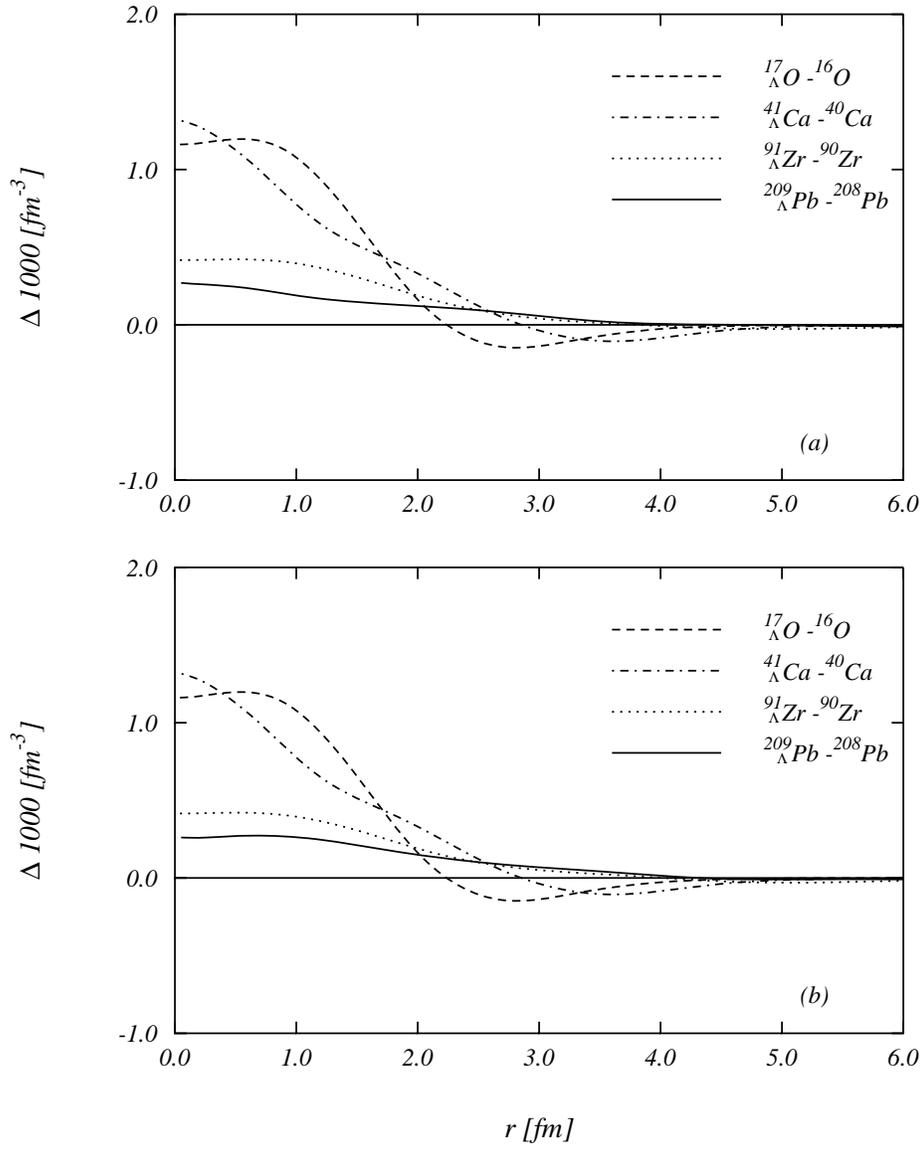}
\end{center}
\vskip 3.5 cm
\caption{ Differences between proton (panel $a$) and 
   neutron (panel $b$) densities with and without core
  polarization. 
 }
\label{fig:core}
\end{figure}

The effects of the $\la$N correlations on the one--body densities are 
shown in Fig.\ref{fig:dnuc}, where the proton densities in four nuclei 
are compared. The dotted lines are the IPM densities, the dashed lines 
are the densities obtained in a purely nucleonic  FHNC calculations, and, 
finally, the full curves represent the proton densities obtained when 
a $\la$ hyperon  is added in the $s$ wave to the doubly magic nucleonic core. 
In general, correlation effects are more important in Oxygen and
Calcium than in  
the other two nuclei. Moreover, the presence of the $\la$ does not heavily  
modify the nucleonic densities. This is better shown in Fig.\ref{fig:core}, 
where the differences between proton (panel $a$) and neutron
(panel $b$) densities with and without the $\la$ are given. 
The differences shown in the figure have been amplified by a factor 1000. 
The heavier nuclei seem to be much more 
stable against deformations produced by the presence of the $\la$.

\begin{figure}
\begin{center}
\vspace*{1.cm}
\hspace*{-2.0 cm}
\leavevmode
\epsfysize = 350pt
\epsfbox[70 200 500 650]{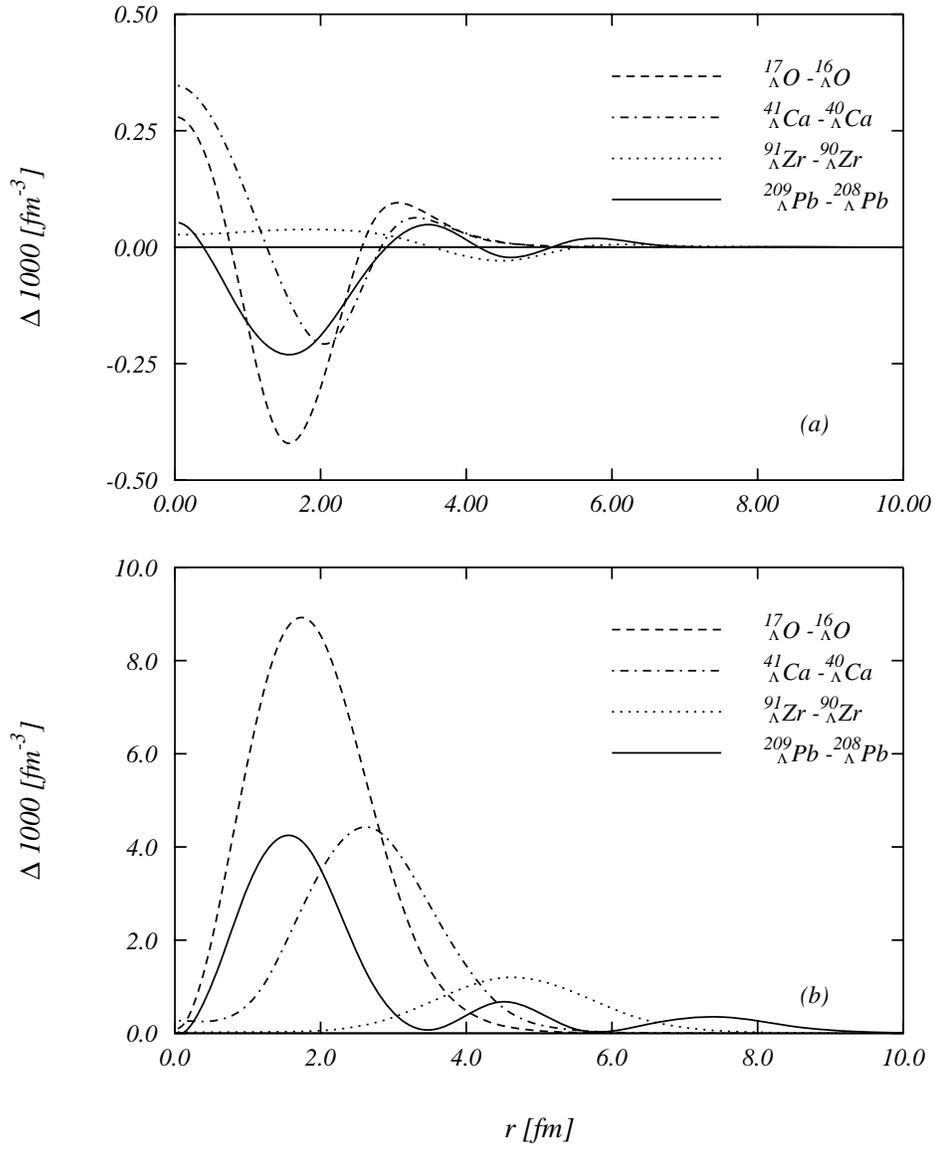}
\end{center}
\vskip 3.5 cm
\caption{ Differences between proton (panel $a$) and 
   neutron (panel $b$) densities 
  of isotopic hypernuclei with A+1 and A hadrons. }
\label{fig:iso}
\end{figure}

This picture might change when the $\la$ is inserted in the nucleonic
core by substituting the last bound neutron. The differences,
multiplied by 1000, between the nucleonic densities of the isotopic
hypernuclei with A+1 and A hadrons are given in Fig.\ref{fig:iso}. The 
panels show the differences between the protons (upper) and neutron 
(lower) densities. The conservation of the charge and mass
numbers for each hypernucleus implies that the curves in the upper
panel are normalized to zero, while those in the lower panel are
normalized to one.

The analysis of this figure indicates that the nucleonic part is more
perturbed if the $\la$ substitutes a neutron having a low angular momentum, 
like in O, Ca, and Pb. In Zr the neutron modified into a
$\la$ is lying on the $1g_{9/2}$ level ($l$=5), and its wave function is
peaked at the surface of the nucleus, as shown in 
panel $(b)$ of  Fig.\ref{fig:iso}. In any case,  
heavy nuclei have again densities more rigid 
against deformations induced by the hyperon.

\begin{figure}
\begin{center}
\vspace*{1.cm}
\hspace*{-2.0 cm}
\leavevmode
\epsfysize = 350pt
\epsfbox[70 200 500 650]{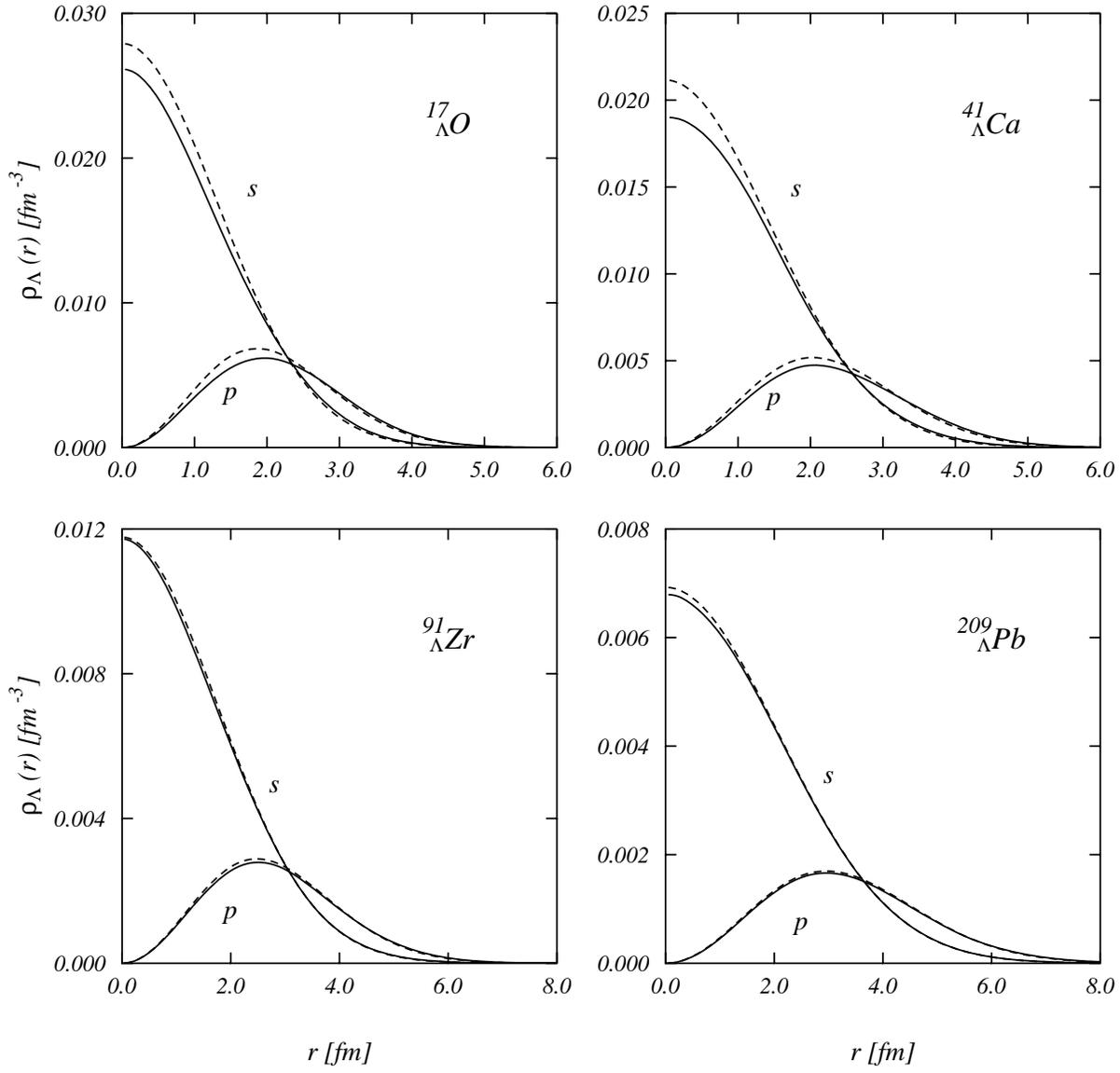}
\end{center}
\vskip 3.5 cm
\caption{ $\la$ densities in $s$ and $p$ states for various hypernuclei.
          Dashed lines: IPM; full line: FHNC results.}
\label{fig:dlam}
\end{figure}

The $\la$ densities in $^{17}_{~\Lambda}$O, $^{41}_{~\Lambda}$Ca,
$^{91}_{~\Lambda}$Zr and  $^{209}_{~~\Lambda}$Pb for $s$ and $p$ waves 
are shown in Fig.\ref{fig:dlam}. The dashed lines are the IPM densities 
and the full ones those obtained in the full calculations. Results very 
similar have been obtained for the $^{16}_{~\Lambda}$O, $^{40}_{~\Lambda}$Ca,
$^{90}_{~\Lambda}$Zr and $^{208}_{~~\Lambda}$Pb hypernuclei. Again,
the differences between the IPM and FHNC approaches are smaller for
the heavier systems.

\section{Conclusions}
\label{conc}
In this work we have studied some properties of 
hypernuclei containing one $\Lambda$ hyperon in the framework of 
the correlated basis functions theory. The $\Lambda$ binding 
energies in the $s$, $p$ and $d$ states, its density and the
rearrangement  part of the one--nucleon 
densities have been computed for hypernuclei 
whose  nucleonic core has a doubly closed shell structure in $jj$ 
coupling. The hyperon has been either added to the 
core ($^{13}_{~\Lambda}$C, $^{17}_{~\Lambda}$O, $^{41}_{~\Lambda}$Ca, 
$^{49}_{~\Lambda}$Ca, $^{91}_{~\Lambda}$Zr and  $^{209}_{~~\Lambda}$Pb) 
or substituted to a neutron in its highest shell 
 ($^{12}_{~\Lambda}$C, $^{16}_{~\Lambda}$O, $^{40}_{~\Lambda}$Ca, 
$^{48}_{~\Lambda}$Ca, $^{90}_{~\Lambda}$Zr and  $^{208}_{~~\Lambda}$Pb). 

The correlated wave function contains gaussian, Jastrow correlations for the 
NN and $\la$N pairs. The correlations 
act on a hypernuclear shell model wave function generated  
 $i)$ by a harmonic oscillator or Woods--Saxon mean field for the nucleonic 
part and $ii)$ by a harmonic oscillator well for the $\Lambda$ single 
particle potential. Cluster expansion and Fermi hypernetted chain resummation 
technique have been used. The energy of the system has been minimized with 
respect to variations on the parameters of the wave function. We have 
employed semirealistic hamiltonians with the S3 two--nucleon potential of 
Afnan and Tang, the Bodmer and Usmani $\Lambda$N potential and 
either a model of $\Lambda$NN three--particle interaction, still proposed 
by Bodmer and Usmani, or a density dependent modification of the $\Lambda$N 
force, in order to take into account many--hadron interactions.

Using the $\la$NN potential with Jastrow correlations severely underestimates 
the $\la$ binding energy, since the important, attractive tensor
component of the  
two--pion exchange potential does not contribute in this model. In contrast, 
the density dependent modification of the $\la$N potential, fitted to the 
$\la$ binding in nuclear matter, provides results for the $\la$ binding 
energy in nuclei in encouraging agreement with the experimental
data. In fact,  
the disagreement is less than 10$\%$ in all hypernuclei, for both $s$ 
and $p$ $\la$ states. The only exception is provided by the carbon
hypernuclei, 
showing a discrepancy with the experiments of $\sim$30$\%$ that could be 
ascribed to the inefficiency of a spherically symmetric description of 
these systems. Moreover, our analysis stresses the importance of adopting a 
good independent particle wave function as a starting point. In fact, 
the Woods--Saxon nucleonic mean field, fitted to the experimental 
one--nucleon densities, gives a better description of the $\la$ binding than 
the harmonic oscillator model, especially in the heavy hypernuclei. 

The nucleonic core polarization effects due to the presence of the 
hyperon are more important in the lighter hypernuclei than in the 
heavier ones, both for the energy and for the one--nucleon densities. 
The rearrangement contributions to the $\la$ binding energy go from 
$\sim 20\%$ in C to $\sim 5\%$ in Pb and nuclear matter. As 
far as the $\la$--density is concerned, the influence of the correlations 
is much more visible in O and Ca than in the heavier Zr and Pb.

The results shown in this paper have been obtained by means of relatively 
simple hamiltonians, which entitle to use equally simple wave functions. 
However, hyperon--nucleon interactions may now be built on more microscopic 
grounds. It is to be expected that an effective use of these
hamiltonians will ask for a more sophisticated correlation, with a
strong dependence on the relative state of the correlated pair. 
The situation closely resembles 
that in doubly closed shell nuclei and in nuclear matter, where the 
introduction of modern potentials has prompted the extension of the 
correlated basis functions and FHNC theories in that direction. 
Another interesting field of application of this technology is the 
study of hypernuclei with two or more hyperons, especially in view 
of a better determination of the hyperon--hyperon interaction. 

\section*{Acknowledgments}
This work has been partially supported by MURST through the  
{\sl Progetto di Ricerca di Interesse Nazionale: 
Fisica teorica del nucleo atomico e dei sistemi a molticorpi} and by
the Spanish Direcci\'on General de Ciencia y Tecnolog\'{\i}a under project
PB98-1318.

\section*{Appendix}

In this Appendix we collect the different contributions to 
the rearrangement part of the one-- and two--body densities, 
where one of the internal particles must be a $\Lambda$ hyperon. 
Following this rule, we obtain:

\begin{eqnarray}
U_{d,\Lambda}^\alpha (r_1) & = & 
\int d \br_2 \Biggl\{ \xi_d^{\Lambda} (r_2) \left(
X_{dd}^{\alpha \Lambda} (\br_1,\br_2)-
E_{dd}^{\alpha \Lambda} (\br_1,\br_2) -
S_{dd}^{\alpha \Lambda} (\br_1,\br_2) 
(g_{dd}^{\alpha \Lambda} (\br_1,\br_2) -1) \right) 
\nonumber \\
& + & \sum_\beta \Biggl( \xi_d^\beta (r_2) \Bigl( 
X_{dd,\Lambda}^{\alpha \beta} (\br_1,\br_2) - 
E_{dd,\Lambda}^{\alpha \beta} (\br_1,\br_2) -
S_{dd,\Lambda}^{\alpha \beta} (\br_1,\br_2)
(g_{dd}^{\alpha \beta} (\br_1,\br_2) -1) \nonumber \\
& & \hspace*{1.6cm} -S_{dd}^{\alpha \beta} (\br_1,\br_2) 
g_{dd,\Lambda}^{\alpha \beta} (\br_1,\br_2) \Bigr)  \\
& & \hspace*{0.5cm} +  \xi_e^\beta (r_2) \Bigl( 
X_{de,\Lambda}^{\alpha \beta} (\br_1,\br_2) - 
E_{de,\Lambda}^{\alpha \beta} (\br_1,\br_2) -
S_{de,\Lambda}^{\alpha \beta} (\br_1,\br_2)
(g_{dd}^{\alpha \beta} (\br_1,\br_2) -1) \nonumber \\
& & \hspace*{1.6cm} 
-S_{de}^{\alpha \beta} (\br_1,\br_2) 
g_{dd,\Lambda}^{\alpha \beta} (\br_1,\br_2) 
-S_{dd,\Lambda}^{\alpha \beta} (\br_1,\br_2) 
g_{de}^{\alpha \beta} (\br_1,\br_2)
\nonumber \\ 
& & \hspace*{1.6cm} 
-S_{dd}^{\alpha \beta} (\br_1,\br_2) 
g_{de,\Lambda}^{\alpha \beta} (\br_1,\br_2) 
\Bigr) \nonumber \\
& & \hspace*{0.5cm} + \xi_{d,\Lambda}^\beta (r_2) \Bigl( 
X_{dd}^{\alpha \beta} (\br_1,\br_2) - 
E_{dd}^{\alpha \beta} (\br_1,\br_2) -
S_{dd}^{\alpha \beta} (\br_1,\br_2)
(g_{dd}^{\alpha \beta} (\br_1,\br_2) -1) \Bigr) \nonumber \\
& & \hspace*{0.5cm} +  \xi_{e,\Lambda}^\beta (r_2) \Bigl( 
X_{de}^{\alpha \beta} (\br_1,\br_2) - 
E_{de}^{\alpha \beta} (\br_1,\br_2) -
S_{de}^{\alpha \beta} (\br_1,\br_2)
(g_{dd}^{\alpha \beta} (\br_1,\br_2) -1) \nonumber \\
& & \hspace*{1.6cm} 
-S_{dd}^{\alpha \beta} (\br_1,\br_2) 
g_{de}^{\alpha \beta} (\br_1,\br_2) 
\Bigr) \Biggr) \Biggr\} ~~, \nonumber \\
U_{e,\Lambda}^\alpha (r_1) & = & \int d \br_2 
\Biggl\{ \xi_d^{\Lambda} (r_1) \Bigl(
X_{ed}^{\alpha \Lambda} (\br_1,\br_2)-
E_{ed}^{\alpha \Lambda} (\br_1,\br_2) -
S_{ed}^{\alpha \Lambda} (\br_1,\br_2) 
(g_{dd}^{\alpha \Lambda} (\br_1,\br_2) -1) \nonumber \\
& & \hspace*{2.1cm} -S_{dd}^{\alpha \Lambda} (\br_1,\br_2) 
g_{ed}^{\alpha \Lambda} (\br_1,\br_2) \Bigr)  \\
& + & \sum_\beta \Biggl( \xi_d^\beta (r_2) \Bigl( 
X_{ed,\Lambda}^{\alpha \beta} (\br_1,\br_2) - 
E_{ed,\Lambda}^{\alpha \beta} (\br_1,\br_2) -
S_{ed,\Lambda}^{\alpha \beta} (\br_1,\br_2)
(g_{dd}^{\alpha \beta} (\br_1,\br_2) -1) \nonumber \\
& & \hspace*{1.6cm} 
-S_{ed}^{\alpha \beta} (\br_1,\br_2) 
g_{dd,\Lambda}^{\alpha \beta} (\br_1,\br_2) 
-S_{dd,\Lambda}^{\alpha \beta} (\br_1,\br_2) 
g_{ed}^{\alpha \beta} (\br_1,\br_2) 
\nonumber \\ & & \hspace*{1.6cm} 
-S_{dd}^{\alpha \beta} (\br_1,\br_2) 
g_{ed,\Lambda}^{\alpha \beta} (\br_1,\br_2) 
\Bigr) \nonumber \\
& & \hspace*{0.5cm} +  \xi_e^\beta (r_2) \Bigl( 
X_{ee,\Lambda}^{\alpha \beta} (\br_1,\br_2) - 
E_{ee,\Lambda}^{\alpha \beta} (\br_1,\br_2) -
S_{ee,\Lambda}^{\alpha \beta} (\br_1,\br_2)
(g_{dd}^{\alpha \beta} (\br_1,\br_2) -1) 
\nonumber \\ & & \hspace*{1.6cm} 
-S_{ee}^{\alpha \beta} (\br_1,\br_2) 
g_{dd,\Lambda}^{\alpha \beta} (\br_1,\br_2) 
-S_{ed,\Lambda}^{\alpha \beta} (\br_1,\br_2) 
g_{de}^{\alpha \beta} (\br_1,\br_2) 
\nonumber \\ & & \hspace*{1.6cm} 
-S_{ed}^{\alpha \beta} (\br_1,\br_2) 
g_{de,\Lambda}^{\alpha \beta} (\br_1,\br_2) 
-S_{de,\Lambda}^{\alpha \beta} (\br_1,\br_2) 
g_{ed}^{\alpha \beta} (\br_1,\br_2) 
\nonumber \\ & & \hspace*{1.6cm} 
-S_{de}^{\alpha \beta} (\br_1,\br_2) 
g_{ed,\Lambda}^{\alpha \beta} (\br_1,\br_2) 
-S_{dd,\Lambda}^{\alpha \beta} (\br_1,\br_2) 
g_{ee}^{\alpha \beta} (\br_1,\br_2) 
 \nonumber \\
& & \hspace*{1.6cm} 
-S_{dd}^{\alpha \beta} (\br_1,\br_2) 
g_{ee,\Lambda}^{\alpha \beta} (\br_1,\br_2) 
\Bigr) \nonumber \\
& & \hspace*{0.5cm} + \xi_{d,\Lambda}^\beta (r_2) \Bigl( 
X_{ed}^{\alpha \beta} (\br_1,\br_2) - 
E_{ed}^{\alpha \beta} (\br_1,\br_2) -
S_{ed}^{\alpha \beta} (\br_1,\br_2)
(g_{dd}^{\alpha \beta} (\br_1,\br_2) -1) 
\nonumber \\ & & \hspace*{1.6cm} 
-S_{dd}^{\alpha \beta} (\br_1,\br_2) 
g_{ed}^{\alpha \beta} (\br_1,\br_2) 
\Bigr) \nonumber \\
& & \hspace*{0.5cm} +  \xi_{e,\Lambda}^\beta (r_2) \Bigl( 
X_{ee}^{\alpha \beta} (\br_1,\br_2) - 
E_{ee}^{\alpha \beta} (\br_1,\br_2) -
S_{ee}^{\alpha \beta} (\br_1,\br_2)
(g_{dd}^{\alpha \beta} (\br_1,\br_2) -1) \nonumber \\
& & \hspace*{1.6cm} 
-S_{ed}^{\alpha \beta} (\br_1,\br_2) 
g_{de}^{\alpha \beta} (\br_1,\br_2) 
-S_{de}^{\alpha \beta} (\br_1,\br_2) 
g_{ed}^{\alpha \beta} (\br_1,\br_2) 
\nonumber \\ & & \hspace*{1.6cm} 
-S_{dd}^{\alpha \beta} (\br_1,\br_2) 
g_{ee}^{\alpha \beta} (\br_1,\br_2) 
\Bigr) \Biggr) \nonumber \\
& + & 4 \xi_e^\alpha (r_2) \Bigl( 
S_{cc,P}^\alpha (\br_1,\br_2) 
g_{cc,P,\Lambda}^\alpha (\br_1,\br_2)+
S_{cc,P,\Lambda}^\alpha (\br_1,\br_2) 
g_{cc,P}^\alpha (\br_1,\br_2)+ \nonumber \\
& & \hspace*{1.2cm}
S_{cc,A}^\alpha (\br_1,\br_2) 
g_{cc,A,\Lambda}^\alpha (\br_1,\br_2)+
S_{cc,A,\Lambda}^\alpha (\br_1,\br_2) 
g_{cc,A}^\alpha (\br_1,\br_2) \Bigr) 
\nonumber \\
& - & 2 \Bigl( 
\rho_{0,P}^\alpha (\br_1,\br_2) 
N_{cc,P,\Lambda}^\alpha (\br_1,\br_2)+
\rho_{0,A}^\alpha (\br_1,\br_2) 
N_{cc,A,\Lambda}^\alpha (\br_1,\br_2) \Bigr)
\nonumber \\
& + & 4 \xi_{e,\Lambda}^\alpha (r_2) \Bigl( 
S_{cc,P}^\alpha (\br_1,\br_2) 
g_{cc,P}^\alpha (\br_1,\br_2)+
S_{cc,A}^\alpha (\br_1,\br_2) 
g_{cc,A}^\alpha (\br_1,\br_2) \Bigr)
\Biggr\}~~.  \nonumber 
\end{eqnarray}

The rearrangement parts of the two--body distribution functions   
are:
\begin{eqnarray}
\rho_{2,1,\Lambda}^{\alpha \beta} (\br_1,\br_2) & = & \xi_d^\alpha (r_1) \Bigl( 
g_{dd,\Lambda}^{\alpha \beta} (\br_1,\br_2 )\xi_d^\beta (r_2)+
g_{dd}^{\alpha \beta} (\br_1,\br_2 )\xi_{d,\Lambda}^\beta (r_2)+ \\
& & \hspace*{1.1cm}
g_{de,\Lambda}^{\alpha \beta} (\br_1,\br_2 )\xi_e^\beta (r_2)+
g_{de}^{\alpha \beta} (\br_1,\br_2 )\xi_{e,\Lambda}^\beta (r_2) \Bigr)  \nonumber \\
& + & \xi_{d,\Lambda}^\alpha (r_1) \Bigl( 
g_{dd}^{\alpha \beta} (\br_1,\br_2 )\xi_d^\beta (r_2)+
g_{de}^{\alpha \beta} (\br_1,\br_2 )\xi_{e}^\beta (r_2) \Bigr)  \nonumber \\
& + & \xi_e^\alpha (r_1) \Bigl( 
g_{ed,\Lambda}^{\alpha \beta} (\br_1,\br_2 )\xi_d^\beta (r_2)+
g_{ed}^{\alpha \beta} (\br_1,\br_2 )\xi_{d,\Lambda}^\beta (r_2)+ \nonumber \\
& & \hspace*{1.1cm}
g_{ee,\Lambda}^{\alpha \beta} (\br_1,\br_2 )\xi_e^\beta (r_2)+
g_{ee}^{\alpha \beta} (\br_1,\br_2 )\xi_{e,\Lambda}^\beta (r_2) \Bigr)  \nonumber \\
& + & \xi_{e,\Lambda}^\alpha (r_1) \Bigl( 
g_{ed}^{\alpha \beta} (\br_1,\br_2 )\xi_d^\beta (r_2)+
g_{ee}^{\alpha \beta} (\br_1,\br_2 )\xi_{e}^\beta (r_2) \Bigr)  \nonumber~~, \\
B_{mn,\Lambda}^{\alpha \beta} (\br_1,\br_2) & = & N_{mn,\Lambda}^{\alpha \beta} (\br_1,\br_2)
+ E_{mn,\Lambda}^{\alpha \beta} (\br_1,\br_2)~~, \\
g_{mn,\Lambda}^{\alpha \beta} (\br_1,\br_2) & = & N_{mn,\Lambda}^{\alpha \beta} (\br_1,\br_2)
+ X_{mn,\Lambda}^{\alpha \beta} (\br_1,\br_2)~~, \\
g_{dd,\Lambda}^{\alpha \beta} (\br_1,\br_2) & = & 
g_{dd}^{\alpha \beta} (\br_1,\br_2) B_{dd,\Lambda}^{\alpha \beta} (\br_1,\br_2)~~, \\ 
g_{de,\Lambda}^{\alpha \beta} (\br_1,\br_2) & = & 
g_{dd}^{\alpha \beta} (\br_1,\br_2) B_{de,\Lambda}^{\alpha \beta} (\br_1,\br_2) +
g_{de}^{\alpha \beta} (\br_1,\br_2) B_{dd,\Lambda}^{\alpha \beta} (\br_1,\br_2)~~, \\ 
g_{ee,\Lambda}^{\alpha \beta} (\br_1,\br_2) & = & 
g_{dd}^{\alpha \beta} (\br_1,\br_2) B_{ee,\Lambda}^{\alpha \beta} (\br_1,\br_2) +
g_{de}^{\alpha \beta} (\br_1,\br_2) B_{ed,\Lambda}^{\alpha \beta} (\br_1,\br_2) \\ 
& + & g_{ed}^{\alpha \beta} (\br_1,\br_2) B_{de,\Lambda}^{\alpha \beta} (\br_1,\br_2)  
+ g_{ee}^{\alpha \beta} (\br_1,\br_2) B_{dd,\Lambda}^{\alpha \beta} (\br_1,\br_2)
\nonumber \\ 
& - & 2 \delta_{\alpha \beta} \Bigl( g_{cc,P}^{\alpha} (\br_1,\br_2) 
B_{cc,P,\Lambda}^{\alpha} (\br_1,\br_2) 
+ g_{cc,A}^{\alpha} (\br_1,\br_2) B_{cc,A,\Lambda}^{\alpha} (\br_1,\br_2) \Bigr)~~, \nonumber \\
g_{cc,D,\Lambda}^{\alpha} (\br_1,\br_2) & = & 
g_{dd}^{\alpha \alpha} (\br_1,\br_2) B_{cc,D,\Lambda}^{\alpha} (\br_1,\br_2) +
g_{cc,D}^{\alpha} (\br_1,\br_2) B_{dd,\Lambda}^{\alpha \alpha} (\br_1,\br_2)~~.   
\end{eqnarray}
The equations for the nodal diagrams are:
\begin{eqnarray}
N_{mn,\Lambda}^{\alpha \beta} (\br_1,\br_2) & = & 
\Bigl( X_{md}^{\alpha \Lambda} (\br_1,\br_3) \xi_d^{\Lambda} (r_3) | 
X_{dn}^{\Lambda \beta} (\br_3,\br_2) +N_{dn}^{\Lambda \beta} (\br_3,\br_2) \Bigr) + \\
& {\displaystyle \sum_{\gamma=p,n} \sum_{l,l'}} \Biggr[ &  
\Bigl( X_{ml,\Lambda}^{\alpha \gamma} (\br_1,\br_3) \xi_{ll'}^{\gamma} (r_3) | 
X_{l'n}^{\gamma \beta} (\br_3,\br_2) +N_{l'n}^{\gamma \beta} (\br_3,\br_2) \Bigr) +
\nonumber \\ 
& & 
\Bigl( X_{ml}^{\alpha \gamma} (\br_1,\br_3) \xi_{ll',\Lambda}^{\gamma} (r_3) | 
X_{l'n}^{\gamma \beta} (\br_3,\br_2) +N_{l'n}^{\gamma \beta} (\br_3,\br_2) \Bigr) + 
\nonumber \\
& & 
\Bigl( X_{ml}^{\alpha \gamma} (\br_1,\br_3) \xi_{ll'}^{\gamma} (r_3) | 
X_{l'n,\Lambda}^{\gamma \beta} (\br_3,\br_2) +N_{l'n,\Lambda}^{\gamma \beta} (\br_3,\br_2) \Bigr) 
\Biggr]~~, \nonumber 
\end{eqnarray} 
with $m,n=d,e$ and 
\begin{eqnarray} 
\xi_{ll'}^\gamma (r_3) & = &  \delta_{dd,ll'} \xi_d^\gamma (r_3)+
(1-\delta_{dd,ll'}) \xi_e^\gamma (r_3)~~, \\
\xi_{ll',\Lambda}^\gamma (r_3) & = &  \delta_{dd,ll'} \xi_{d,\Lambda}^\gamma (r_3)+
(1-\delta_{dd,ll'}) \xi_{e,\Lambda}^\gamma (r_3)~~. 
\end{eqnarray}
Finally, the equations for the $cc$ nodal are:
\begin{eqnarray}
N_{cc,D,\Lambda}^{(x) \alpha} (\br_1,\br_2) & = & 
\Bigl( X_{cc,D,\Lambda}^{\alpha} (\br_1,\br_3) \xi_e^{\alpha} (r_3) | 
g_{cc,P}^{\alpha} (\br_3,\br_2) \Bigr) + \\
& & \Bigl( X_{cc,D}^{\alpha} (\br_1,\br_3) \xi_{e,\Lambda}^{\alpha} (r_3) | 
g_{cc,P}^{\alpha} (\br_3,\br_2) \Bigr) + \nonumber  \\
& & \Bigl( X_{cc,D}^{\alpha} (\br_1,\br_3) \xi_e^{\alpha} (r_3) | 
 g_{cc,P,\Lambda}^{\alpha} (\br_3,\br_2) \Bigr) +\nonumber \\  
&  s_D \Bigl[ &  \Bigl( X_{cc,E,\Lambda}^{\alpha} (\br_1,\br_3) \xi_e^{\alpha} (r_3) | 
g_{cc,A}^{\alpha} (\br_3,\br_2) \Bigr) + \nonumber \\
& & \Bigl( X_{cc,E}^{\alpha} (\br_1,\br_3) \xi_{e,\Lambda}^{\alpha} (r_3) | 
g_{cc,A}^{\alpha} (\br_3,\br_2) \Bigr) +  \nonumber \\
& & \Bigl( X_{cc,E}^{\alpha} (\br_1,\br_3) \xi_e^{\alpha} (r_3) | 
g_{cc,A,\Lambda}^{\alpha} (\br_3,\br_2) \Bigr) \ \ \ \Bigr] \nonumber \\
-N_{cc,D,\Lambda}^{(\rho) \alpha} (\br_1,\br_2) & = &
 \Bigl( \rho_{D}^{\alpha} (\br_1,\br_3) \xi_{e,\Lambda}^{\alpha} (r_3) | 
g_{cc,P}^{\alpha} (\br_3,\br_2) \Bigr)+    \\
& & \Bigl( \rho_{D}^{\alpha} (\br_1,\br_3) \xi_{e}^{\alpha} (r_3) | 
X_{cc,P,\Lambda}^{\alpha} (\br_3,\br_2)+N_{cc,P,\Lambda}^{(x) \alpha} (\br_3,\br_2) \Bigr)+ 
\nonumber  \\
& & \Bigl( \rho_{D}^{\alpha} (\br_1,\br_3) (\xi_e^{\alpha} (r_3)-1) | 
 N_{cc,P,\Lambda}^{(\rho) \alpha} (\br_3,\br_2) \Bigr)+
\nonumber \\
&  s_D \Bigl[ &
\Bigl( \rho_{E}^{\alpha} (\br_1,\br_3) \xi_{e,\Lambda}^{\alpha} (r_3) | 
g_{cc,A}^{\alpha} (\br_3,\br_2) \Bigr) + \nonumber \\
& & \Bigl( \rho_{E}^{\alpha} (\br_1,\br_3) \xi_{e}^{\alpha} (r_3) | 
X_{cc,A,\Lambda}^{\alpha} (\br_3,\br_2)+N_{cc,A,\Lambda}^{(x) \alpha} (\br_3,\br_2) \Bigr) +
\nonumber  \\
& & \Bigl( \rho_{E}^{\alpha} (\br_1,\br_3) (\xi_e^{\alpha} (r_3)-1) | 
 N_{cc,A,\Lambda}^{(\rho) \alpha} (\br_3,\br_2) \Bigr) \ \ \ \Bigr]~~, \nonumber
\end{eqnarray} 
where $D=P,A$. If $D=P$, then $E=A$ and $s_P=-1$; 
if $D=A$, then $E=P$ and $s_A=1$.


\newpage
%
%
\begin{table}[h]
\begin{center}
\begin{tabular}{l|rrr|rr|rr}
\hline
            &\multicolumn{3}{c|}{HO}  
            &\multicolumn{2}{c|}{WS} 
            &\multicolumn{2}{c}{exp} \\
\hline
           &$b_N$ & $R $ & $B/A$  & $R$  & $B/A$   & $R$  & $B/A$ \\
\hline
$^{12}$C   & 1.56 & 2.48 & 2.63 & 2.47  & 2.78 & 2.47 & 7.680 \\
$^{16}$O   & 1.48 & 2.43 & 7.07 & 2.69  & 6.29 & 2.73 & 7.976 \\
$^{40}$Ca  & 1.66 & 3.08 & 9.00 & 3.29  & 8.12 & 3.48 & 8.551 \\
$^{48}$Ca  & 1.62 & 3.00 & 7.57 & 3.35  & 6.79 & 3.48 & 8.666 \\
$^{90}$Zr  & 1.74 & 3.57 & 10.07 & 4.09 & 7.30 & 4.04 & 8.710 \\
$^{208}$Pb & 2.05 & 4.67 & 10.24 & 5.52 & 8.03 & 5.50 & 7.867 \\
\hline
\end{tabular}
\end{center}
\caption{Optimum values of the nucleonic oscillator constant, 
         $b_N$ (in fm), of the rms radii $R$ (in fm)   
         and of the binding energy per particle $B/A$ (in MeV) for
         different nuclei. The Woods-Saxon potential parameters are
         those of Ref. \protect\cite{ari96}.}
\label{tab:oscc}
\end{table}
%
%
\begin{table}[h]
\begin{center}
\begin{tabular}{l|rrr|rrr}
\hline
            &\multicolumn{3}{c|}{HO}  
            &\multicolumn{3}{c}{WS} \\
\hline
 & $b_{\Lambda}$  & $\alpha_{\Lambda}$  & $\beta_{\Lambda}$
 & $b_{\Lambda}$  & $\alpha_{\Lambda}$  & $\beta_{\Lambda}$ \\
\hline
$^{13}_{~\Lambda}$C    & 1.92 & 0.80 & 3.0  & 1.92 & 0.80 & 3.1  \\
$^{17}_{~\Lambda}$O    & 1.86 & 0.80 & 2.9  & 1.86 & 0.80 & 2.9  \\
$^{41}_{~\Lambda}$Ca   & 2.00 & 0.75 & 2.9  & 2.04 & 0.75 & 2.9  \\
$^{49}_{~\Lambda}$Ca   & 2.00 & 0.75 & 2.9  & 2.04 & 0.75 & 3.0  \\
$^{91}_{~\Lambda}$Zr   & 2.24 & 0.70 & 2.9  & 2.48 & 0.80 & 3.1  \\
$^{209}_{~~\Lambda}$Pb & 2.68 & 0.70 & 3.0  & 2.96 & 0.80 & 3.1  \\
\hline
\end{tabular}
\end{center}
\caption{Optimum values of the $\Lambda$ oscillator constant, $b_{\Lambda}$ 
(in fm), and of the parameters of $f_\Lambda(r)$, $\alpha_\Lambda$ and 
$\beta_\Lambda$ (in fm$^{-2}$) for the two nuclear mean fields used in this
work. In nuclear matter,  $\alpha_{\Lambda}$=0.70 and 
$\beta_{\Lambda}$=3.2 fm$^{-2}$.
}
\label{tab:osck}
\end{table}
%
%
\begin{table}[t]
\begin{center}
\begin{tabular}{l|rrrrrr}
\hline
Hypernucleus & $^{17}_{~\Lambda}$O & $^{41}_{~\Lambda}$Ca
&$^{49}_{~\Lambda}$Ca & 
$^{91}_{~\Lambda}$Zr & $^{209}_{~~\Lambda}$Pb & NM \\ 
\hline
$T_{\Lambda}^I$ & 17.12 & 18.98 & 20.78 & 19.96 & 19.24 & 10.13 \\ 
$T_{\Lambda}^R$ & -1.66 & -2.65 & -3.63 & -3.87 & -3.52 & -0.49 \\
$T_{\Lambda}^{cm}$ & 0.46 & 0.14 & 0.13 & 0.06 & 0.02 \\
$V_{\Lambda}^I$ & -42.76 & -61.63 & -69.78 & -80.54 & -87.87 & -59.89 \\
$V_{\Lambda}^R$ & 2.31 & 3.59 & 4.82 & 5.08 & 5.09 & 2.22 \\
$B^{(2)}_{\Lambda}$ & 24.53 & 41.57 & 47.68 & 59.31 & 67.04 & 48.03 \\
\hline
$V_{\Lambda}^3$ & 17.72 & 37.80 & 49.12 & 74.41 & 77.72 \\
$B^{(3)}_{\Lambda}$ & 6.81 & 3.77 & -1.44 & -15.10 & -10.68 \\
\hline
$\Delta V^{DDP}_{\Lambda}$ & 12.75 & 21.81 & 26.99 & 34.65 & 37.89 & 17.77 \\
$B^{DDP}_{\Lambda}$ & 11.78 & 19.76 & 20.69 & 24.66 & 29.15 & 30.26 \\
\hline
\end{tabular}
\end{center}
\caption{Contributions, in MeV, to the $\Lambda$ binding energy for various 
         hypernuclei calculated within the HO model and compared with the
         nuclear matter (NM) results. See text.}
\label{tab:detail}
\end{table}

\begin{table}
\begin{center}
\begin{tabular}{l|rrr}
\hline
             & $B_{\Lambda}^I$ & $B_{\Lambda}^R$ & $B_{\Lambda}$ \\ 
\hline
$^{12}_{~\Lambda}$C   &   9.54 &  -1.85 &   7.69 \\
$^{13}_{~\Lambda}$C   &  10.16 &  -1.74 &   8.42  \\
$^{16}_{~\Lambda}$O   &  12.41 &  -1.39 &  11.02 \\
$^{17}_{~\Lambda}$O   &  12.89 &  -1.11 &  11.78  \\
$^{40}_{~\Lambda}$Ca  &  20.76 &  -1.18 &  19.58 \\
$^{41}_{~\Lambda}$Ca  &  20.84 &  -1.08 &  19.76  \\
$^{48}_{~\Lambda}$Ca  &  21.90 &  -1.39 &  20.51  \\
$^{49}_{~\Lambda}$Ca  &  22.01 &  -1.32 &  20.69  \\
$^{90}_{~\Lambda}$Zr  &  25.88 &  -1.31 &  24.57  \\
$^{91}_{~\Lambda}$Zr  &  25.93 &  -1.27 &  24.66  \\
$^{208}_{~~\Lambda}$Pb &  30.72 &  -1.60 &  29.12 \\
$^{209}_{~~\Lambda}$Pb &  30.74 &  -1.59 &  29.15  \\
NM                   &  31.99 &  -1.73 &  30.26   \\
\hline
\end{tabular}
\end{center}
\caption{ Interaction, rearrangement and total binding energies, in MeV, 
          of a $\Lambda$ in the $1s$ state with the HO nucleonic mean
          field. Also the energy of a single $\la$ in 
          nuclear matter is shown.} 
\label{tab:sho}
\end{table}

\begin{table}
\begin{center}
\begin{tabular}{l|rr}
\hline
              & $B_{\Lambda}(1p)$ &  $B_{\Lambda}(1d)$  \\
\hline
$^{16}_{~\Lambda}$O   &   0.73  & \\
$^{17}_{~\Lambda}$O   &   1.38  & \\
$^{40}_{~\Lambda}$Ca  &  10.01  & 0.85 \\
$^{41}_{~\Lambda}$Ca  &  10.32  & 1.13 \\
$^{48}_{~\Lambda}$Ca  &  11.08  & 2.00 \\
$^{49}_{~\Lambda}$Ca  &  11.27  & 2.19 \\
$^{90}_{~\Lambda}$Zr  &  16.85  & 8.69 \\
$^{91}_{~\Lambda}$Zr  &  16.95  & 8.80 \\
$^{208}_{~~\Lambda}$Pb &  23.54 & 17.18  \\
$^{209}_{~~\Lambda}$Pb &  23.58 & 17.24  \\
\hline
\end{tabular}
\end{center}
\caption{$\Lambda$ binding energies, in MeV, in the $1p$ and $1d$  state
  calculated with HO nucleonic mean field.}
\label{tab:pho}
\end{table}

\begin{table}
\begin{center}
\begin{tabular}{l|rr|rr|rr}
\hline
          & \multicolumn{2}{c|} {$B_{\Lambda}(1s)$} 
         &\multicolumn{2}{c|} {$B_{\Lambda}(1p)$} 
         &\multicolumn{2}{c} {$B_{\Lambda}(1d)$}  \\ 
         & th & exp & th & exp & th & exp  \\
\hline
$^{12}_{~\Lambda}$C   &  7.57 & 10.8$\pm$0.1  &    & 0.1$\pm$0.5 &   & \\
$^{13}_{~\Lambda}$C   &  8.30 & 11.7$\pm$0.1 &    & 0.8$\pm$0.5 &   & \\
$^{16}_{~\Lambda}$O   &  11.21 & 12.5$\pm$0.4 & 1.12 & 2.5$\pm$0.4  & &  \\
$^{17}_{~\Lambda}$O   &  11.99 & & 1.76   & & &  \\
$^{40}_{~\Lambda}$Ca  &  19.61 & 18.7$\pm$1.1 & 10.27 & 10.1$\pm$0.3 
                                              &  1.37 & 1.0$\pm$0.5 \\
$^{41}_{~\Lambda}$Ca  &  19.96 & & 10.61  & & 1.64 &  \\
$^{48}_{~\Lambda}$Ca  &  21.14 & & 11.89  & & 2.98  & \\
$^{49}_{~\Lambda}$Ca  &  21.34 & & 12.10  & & 3.19  &\\
$^{90}_{~\Lambda}$Zr  &  23.19 & 22.1$\pm$1.6$^*$ 
                      & 16.76  & 16.0$\pm$1.0$^*$ 
                      & 9.96   & 9.5$\pm$1.0$^*$\\
$^{91}_{~\Lambda}$Zr  &  23.27 &  & 16.87  & & 10.08 & \\
$^{208}_{~~\Lambda}$Pb &  27.52 & 26.5$\pm$0.5  & 22.78 & 21.3$\pm$0.7 
                                                & 17.33 &  16.5$\pm$0.5 \\
$^{209}_{~~\Lambda}$Pb &  27.55 & & 22.82 & & 17.39 & \\
\hline
\end{tabular}
\end{center}
\caption{$B_\Lambda$, in MeV, for a $\Lambda$ in the $1s$ and $1p$
  states calculated with the WS nucleonic mean fields. The
  experimental binding energies are from Refs. 
  \protect\cite{has96,pil91,may97}. The energies labelled with an
  asterisk have been measured for the $^{89}_{\la}$Y nucleus.}
\label{tab:ws}
\end{table}

\newpage
%
\newpage
\newpage
\newpage
\newpage
\newpage

\end{document}